\documentclass[%
reprint,
superscriptaddress,
amsmath,amssymb,
aps,
prb,
floatfix,
]{revtex4-2}

\usepackage{float}
\usepackage{graphicx}
\usepackage{booktabs}  
\usepackage{multirow}  
\usepackage{siunitx}   
\usepackage{dcolumn}
\usepackage{bm}
\usepackage{hyperref}
\hypersetup{
  colorlinks=true,
  linkcolor=blue,
  urlcolor=blue,
  filecolor=blue,
  urlcolor=blue,
  citecolor=blue,
}
\usepackage{wrapfig}

\usepackage{graphicx}

\usepackage{svg}
\usepackage{mathptmx}
\usepackage{booktabs}

\begin{document}

\preprint{APS/123-QED}

\title{ Self-Optimizing Machine Learning Potential Assisted Automated Workflow for Highly Efficient Complex Systems Material Design}

\author{Jiaxiang Li}
\affiliation{Key Laboratory of Material Simulation Methods \& Software of Ministry of Education, College of Physics, Jilin University, Changchun 130012, China}

\author{Junwei Feng}
\affiliation{Research Center for Crystal Materials, State Key Laboratory of Functional Materials
and Devices for Special Environmental Conditions, Xinjiang Key Laboratory of
Functional Crystal Materials, Xinjiang Technical Institute of Physics and Chemistry,
Chinese Academy of Sciences, 40-1 South Beijing Road, Urumqi 830011, China}

\author{Jie Luo}
\affiliation{Key Laboratory of Material Simulation Methods \& Software of Ministry of Education, College of Physics, Jilin University, Changchun 130012, China}
\affiliation{State Key Laboratory of Superhard Materials, College of Physics, Jilin University, Changchun 130012, China}

\author{Bowen Jiang}
\affiliation{Key Laboratory of Material Simulation Methods \& Software of Ministry of Education, College of Physics, Jilin University, Changchun 130012, China}

\author{Xiangyu Zheng}
\affiliation{Key Laboratory of Material Simulation Methods \& Software of Ministry of Education, College of Physics, Jilin University, Changchun 130012, China}

\author{Qigang Song}
\affiliation{Key Laboratory of Material Simulation Methods \& Software of Ministry of Education, College of Physics, Jilin University, Changchun 130012, China}

\author{Jian Lv}
\affiliation{Key Laboratory of Material Simulation Methods \& Software of Ministry of Education, College of Physics, Jilin University, Changchun 130012, China}

\author{Keith Butler}
\affiliation{Department of Chemistry, University College London, Gordon Street, London, WC1H 0AJ, UK}

\author{Hanyu Liu}
\email{hanyuliu@jlu.edu.cn}
\affiliation{Key Laboratory of Material Simulation Methods \& Software of Ministry of Education, College of Physics, Jilin University, Changchun 130012, China}
\affiliation{State Key Laboratory of Superhard Materials, College of Physics, Jilin University, Changchun 130012, China}
\affiliation{International Center of Future Science, Jilin University, Changchun 130012, China}

\author{Congwei Xie}
\email{cwxie@ms.xjb.ac.cn}
\affiliation{Research Center for Crystal Materials, State Key Laboratory of Functional Materials
and Devices for Special Environmental Conditions, Xinjiang Key Laboratory of
Functional Crystal Materials, Xinjiang Technical Institute of Physics and Chemistry,
Chinese Academy of Sciences, 40-1 South Beijing Road, Urumqi 830011, China}

\author{Yu Xie}
\email{xieyu@jlu.edu.cn}
\affiliation{Key Laboratory of Material Simulation Methods \& Software of Ministry of Education, College of Physics, Jilin University, Changchun 130012, China}
\affiliation{Key Laboratory of Physics and Technology for Advanced Batteries of Ministry of Education, College of Physics, Jilin University, Changchun, 130012, China}
 
\author{Yanming Ma}
\affiliation{Key Laboratory of Material Simulation Methods \& Software of Ministry of Education, College of Physics, Jilin University, Changchun 130012, China}
\affiliation{State Key Laboratory of Superhard Materials, College of Physics, Jilin University, Changchun 130012, China}
\affiliation{International Center of Future Science, Jilin University, Changchun 130012, China}

\date{\today}

\begin{abstract}
Machine learning interatomic potentials have revolutionized complex materials design by enabling rapid exploration of material configurational spaces via crystal structure prediction with \textit{ab initio} accuracy. However, critical challenges persist in ensuring robust generalization to unknown structures and minimizing the requirement for substantial expert knowledge and time-consuming manual interventions.
Here, we propose an automated crystal structure prediction framework built upon the attention-coupled neural networ‌‌k potential to address these limitations. The generalizability of the potential is achieved by sampling regions across the local minima of the potential energy surface, where the self-evolving pipeline‌ autonomously refines the potential iteratively while minimizing human intervention.
The workflow is validated on Mg–Ca–H ternary and Be–P–N–O quaternary systems by exploring nearly 10 million configurations, demonstrating substantial speedup compared to first-principles calculations.  These results underscore the effectiveness of our approach in accelerating the exploration and discovery of complex multi-component functional materials.

\end{abstract}

\maketitle

\section{\label{sec:level1}Introduction}

Functional materials with desired physical and chemical properties are essential for advancing cutting-edge technologies and driving breakthroughs across various disciplines\cite{ceder1998identification, liu2020two, norskov2009towards, de2021materials}. Of particular interest are multi-component materials, whose vast compositional diversity and intricate structural configurations enable the fine-tuning of various properties, such as physical, chemical, and mechanical, offering significant opportunities for discovering high-performance functional materials\cite{walsh2015quest, lu2021noble}.
This tunability is vividly demonstrated in high-$T_\mathrm{c}$ superconductors\cite{bednorz1986possible, kamihara2008iron, song2023stoichiometric}, high strength alloys\cite{otto2013influences}  , fast ionic conducting solid electrolytes\cite{mo2012first}, and wavelength specific nonlinear optical materials\cite{mei1993nonlinear, wu1996linear, xie2023prediction}. However, the enormous configurational space also poses substantial challenges for rational design. Estimates suggest that the inorganic materials space encompasses nearly 10$^{10}$ possible compounds across binary to quaternary systems, whereas conventional experimental approaches to date have accessed only a minute fraction of these compositions and structures\cite{park2025mapping}.
This fundamental limitation severely constrains the systematic exploration and optimization of complex material systems, highlighting the urgent need for advanced design strategies to accelerate materials discovery.

\begin{figure*}[t]
    \includegraphics[width=1.5\columnwidth]{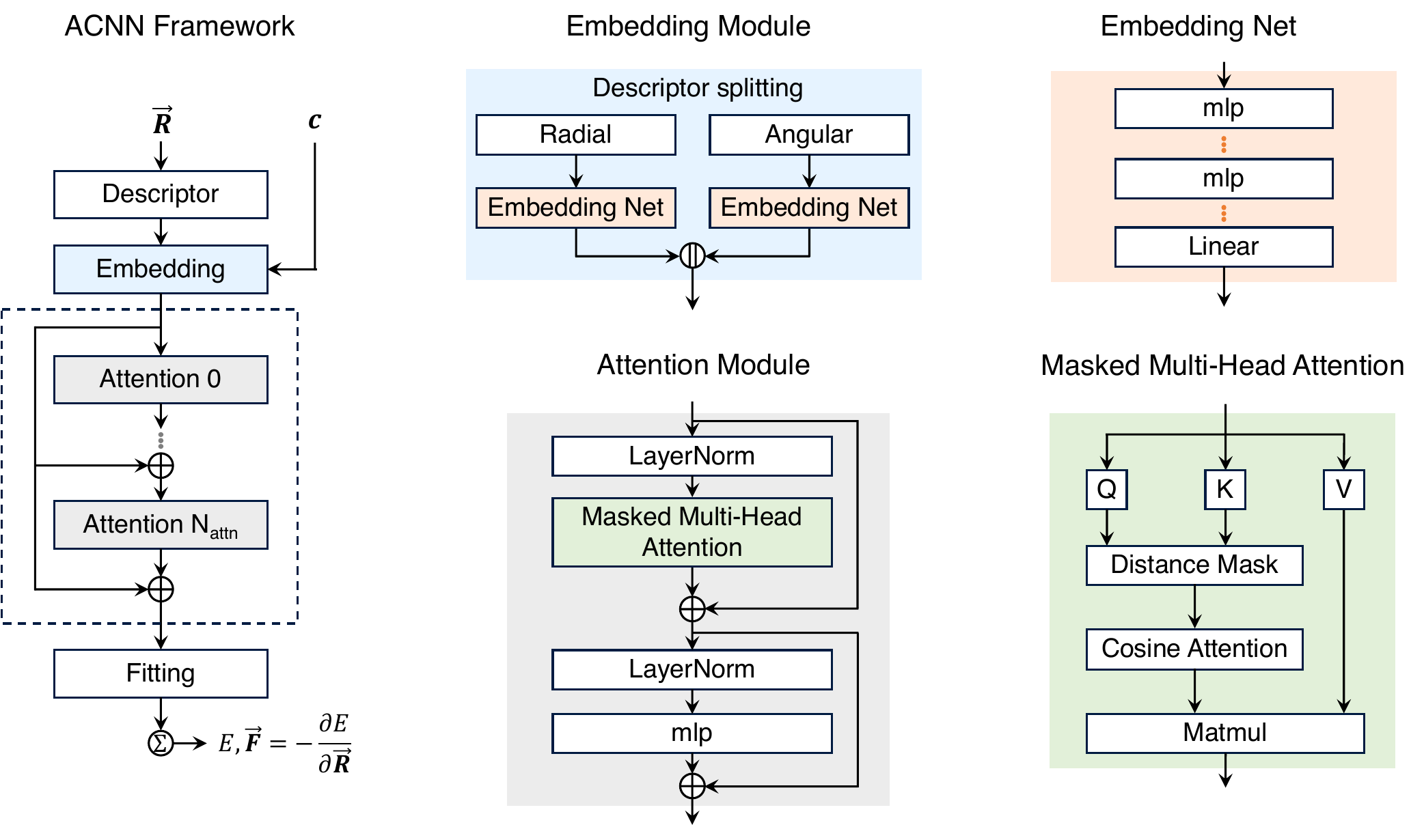}
    \centering
    \caption{ \textbf{Schematic illustration of the ACNN architecture and its modular decomposition.}
    The ACNN architecture consists of four modules: a descriptor, an elemental embedding, a set of attention modules, and an mlp-based fitting module. The symbol $\oplus$ represents Hadamard (element-wise) addition, and $||$ denotes vector concatenation. The multi-head attention modules marked by the dashed box are optional, enabling the ACNN architecture to adaptively switch between different network configurations according to the needs of specific simulation tasks. 
    }
    \label{figure-flow}
\end{figure*}

Over the past two decades, advances in computational approaches have substantially accelerated the discovery of multi-component functional materials. Crystal structure prediction (CSP) methods, such as random sampling (AIRSS)\cite{Pickard_2011}, particle swarm optimization (CALYPSO)\cite{wang2010crystal, wang2012calypso}, and evolutionary algorithms (USPEX)\cite{oganov2006crystal}, have become the most widely adopted strategies for materials design without prior knowledge. When combined with density functional theory (DFT), CSP can accurately construct potential energy surfaces (PES) and guide the search for global and local structures of given compositions. Although numerous functional materials have been synthesized under the guidance of DFT-based CSP, the cubic scaling of DFT calculations causes computational demands to increase sharply with system size and complexity, thereby limiting the large-scale application of CSP. Further development of more efficient methods is thus essential to fully unlock the potential of CSP.






The recently emerged neural network–based machine learning interatomic potentials (MLIPs) can deliver accurate PESs comparable to DFT and with much higher transferability, while their computational cost is orders of magnitude lower\cite{behler2007generalized, shapeev2016moment, bartok2010gaussian, zhang2018end}. This offers an efficient way to accelerate CSP methods, enabling the exploration of large and complex materials spaces. A straightforward way to cooperate MLIP with CSP is to use the universal models (uMLIP)\cite{chen2022universal, batatia2022mace}, which are trained on millions of DFT calculations and exhibit a measure of generalizability. However, owing to the inherently interpolative nature of neural networks, uMLIP is also fixed to the provided chemical spaces and bonding environments. Since CSP targets the prediction of previously unknown structures, the extent to which uMLIPs can reliably capture untrained structural features of multi-component complex systems remains debated\cite{juelsholt2025continued}. 
A more sophisticated and widely adopted strategy is to first train an expert model (eMLIP) through an active learning scheme by constructing a representative DFT dataset, and subsequently perform eMLIP-accelerated CSP for materials design. This framework has been successfully applied across a wide range of systems\cite{wang2024concurrent, tong2018accelerating}. Although specifically trained eMLIPs generally exhibit higher reliability and consistency than uMLIPs, their transferability heavily depends on the quality of the representative DFT dataset\cite{tong2020combining}. For complex systems, the number of relevant local minima increases exponentially\cite{Pickard_2011}, making it challenging to assess the adequacy of the actively generated DFT dataset.  Consequently, the resulting CSP may encounter unphysical local minima/maxima that undermine the reliability of structure prediction and potentially obscure meaningful configurations\cite{tong2020combining}. Except for accuracy, the computational efficiency of MLIPs also becomes critical for complex materials design due to the underlying millions of structural optimizations during the CSP process.


Moreover, beyond the difficulty of generating high-quality DFT datasets for robust and efficient MLIPs, the user’s experience in handling MLIPs' training procedure and CSP simulation pipeline exerts a substantial influence on the overall effectiveness of CSP-driven materials design. Fortunately, automation approaches show the ability to solve such a challenge, as demonstrated in the development of DFT-driven high-throughput materials simulation methods\cite{curtarolo2012aflow, pizzi2016aiida}. Therefore, a fully automated workflow combining highly generalized MILPs and corresponding CSP simulations for multi-component materials design is highly desired.

In this work, we present an automated AI-assisted framework for predicting structures of complex systems, integrating MLIP training and refinement, structure screening, anomaly handling, and large-scale optimization on massively parallel computing environments. To this end, we first introduce an expert MLIP, the attention-coupled neural network (ACNN), which achieves sufficient accuracy across broad stoichiometric ranges while maintaining high computational efficiency for high-throughput structural optimization. Based on ACNN, we develop a self-optimizing process that orchestrates the refinement of the ACNN and the CSP procedure, ensuring progressive improvement in model generalizability and discovery of new structures. We validate our approach by exploring Mg–Ca–H ternary and Be–P–N–O quaternary systems, where several new compounds are identified. Compared with DFT, our framework achieves a speedup of four orders of magnitude in both the Mg–Ca–H and Be–P–N–O systems, underscoring its effectiveness and broad applicability.
We anticipate that this framework will help promote the widespread adoption of MLIPs-driven CSP for complex functional materials design and discovery in the coming years.

\begin{figure*}[t]
    \includegraphics[width=2.05\columnwidth]{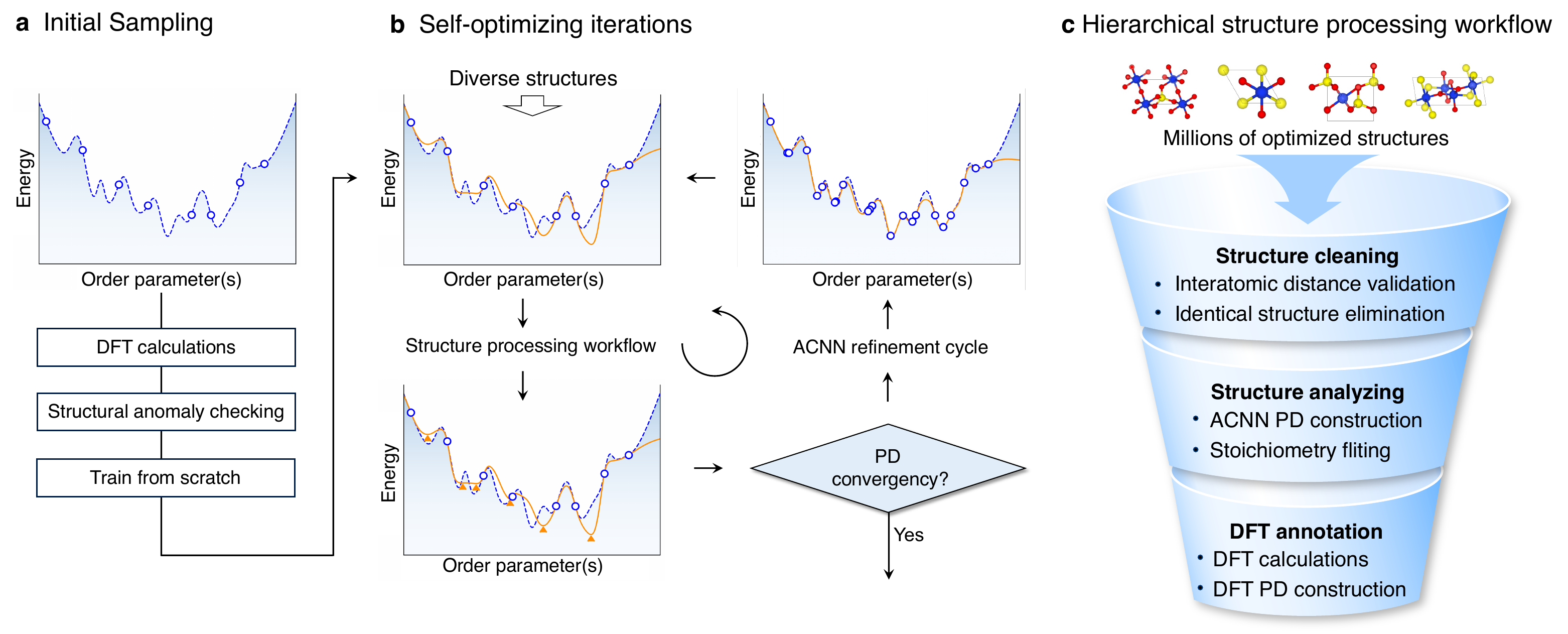}
    \centering
    \caption{\textbf{Automatic iterative self-optimizing schedule}
    \textbf{a} Initial sampling of the PES, which are evaluated by DFT calculations and filtered through structural anomaly checking before training an initial ACNN model.
    \textbf{b} The ACNN self-optimizing approach iteratively models local energy minima, corresponding to stable or metastable phases, with refined predictions (orange solid lines) progressively approaching the underlying PES (blue dashed line). The cycle continues until phase diagram convergence.
    \textbf{c} Hierarchical structure processing workflow applied to millions of optimized structures, which are systematically analyzed and integrated to update the phase diagram.
    }
    \label{figure-main}
\end{figure*}

\section{Results}

\subsection{ACNN architecture}

ACNN incorporates various features extensively discussed in previous studies\cite{behler2007generalized,  bartok2010gaussian, zhang2018end, batzner20223, yu2024systematic}. In particular, it explicitly describes intrinsic translational, rotational, and permutational invariance for scalar quantities (energy), and rotational equivariance for tensorial properties (atomic forces and stress tensors). Additionally, the consistently continuous and smooth differentiability should also be satisfied.
Following \textcite{behler2007generalized}, the total potential energy $E$ is expressed as a sum of atomic contributions $E_i$. 
\begin{equation}
E = \sum_{i}^N E_i,
\end{equation}
where $N$ denotes the total number of atoms.
The force and virial tensors are directly computed from $E$ via
\begin{equation}
\boldsymbol{F}_{i} = - \nabla_i E, \text{ and }  \boldsymbol{W}_i =  \sum_{j \neq i} \boldsymbol{r}_{ij} \otimes 
\frac{\partial E_i }{ \partial \boldsymbol{r}_{ji}}.
\end{equation}
The central term $E_i$ is uniquely defined by the local environment $\bm{R}_i$ of atom $i$,
\begin{equation}
E_i(\boldsymbol{R}_i) = E ( \boldsymbol{r}_{i1}, \boldsymbol{r}_{i2}, ..., \boldsymbol{r}_{iN}),
\end{equation}
in which $\boldsymbol{r_{ij}} = \boldsymbol{r_j} - \boldsymbol{r_i}$ denotes the relative position vector from atom $i$ to atom $j$. ACNN utilizes an analytical descriptor of $\boldsymbol{R_i}$, whose form is derived from the atomic cluster expansion (ACE)\cite{drautz2019atomic, fan2022gpumd} framework. Specifically, the atomic energy $E_i$ is expanded in terms of $\nu$-body correlation functions,
\begin{equation}
\Phi^{(\mathbf{\nu})}(\boldsymbol{r}_{ij_1},\ldots,\boldsymbol{r}_{ij_\nu}).
\end{equation}
The energy is obtained through the nonlinear combination effects of the neural network, expressed as
\begin{equation}
E_i(\boldsymbol{R_i})=\sum_{\nu=1}^{\nu_{\max}} \sum_{\mathbf{\nu}} {\mathbf{c}}_{\mathbf{\nu}}\sum_{j_1,\ldots,j_\nu}\Phi^{(\mathbf{\nu})}(\boldsymbol{r}_{ij_1},\ldots,\boldsymbol{r}_{ij_\nu}) , \label{eq:site_ene}
\end{equation}
where $\boldsymbol{c}_\nu$ are model parameters associated with the neural network architecture. Typically, the interaction is limited to a predefined cutoff sphere with radius $r_c$. To smooth the behavior of atoms at the boundary, an envelope function is defined as
\begin{equation}
f_c(r_{ij})=
\begin{cases}
\frac{1}{2}\Big[1+\cos\Big(\pi\frac{r_{ij}}{r_c}\Big)\Big]&r_{ij}\leq r_c\\0&r_{ij} > r_c
\end{cases}.
\end{equation}
For a central atom $i$, the first-order term, which accounts for pairwise interactions, is given by
\begin{equation}
\Phi_n^{(1)}=\sum_{j\neq i}g_n(r_{ij}),
\end{equation}
where $g_n(r_{ij})$ is a function of the distance, defined as
\begin{equation}
g_n(r_{ij})=\frac{1}{2}\biggl[T_n\left(2\left(\frac{r_{ij}}{r_c}-1\right)^2-1\right)+1\biggr]f_c(r_{ij}),
\end{equation}
with $T_n$ denoteing the Chebyshev polynomial of the first kind of order $n$. The second-order term, which incorporates angular dependencies, is given by
\begin{equation}
\Phi_{nl}^{(2)}=\frac{2l+1}{4\pi}\sum_{j\neq i}\sum_{k\neq i}g_n(r_{ij})g_n(r_{ik})P_l(\cos\theta_{ijk}),
\end{equation}
where $P_l(\cos\theta_{ijk})$ is the Legendre polynomial of order $l$, and $\theta_{ijk}$ is the angle between the vectors $\boldsymbol{r}_{ij}$ and $\boldsymbol{r}_{ik}$.
In general, incorporating higher-order terms improves the completeness of the local environment description and thereby enhances accuracy. However, achieving convergence becomes progressively slower, as evidenced by the tests provided in \textcite{drautz2019atomic}. To balance computational efficiency and precision, ACNN explicitly incorporates up to 3-body interactions, while delegating the remaining residual errors to the nonlinear neural network for further correction.

The overall ACNN architecture consists of three main components: an embedding module, an optional stack of attention modules, and a fitting module, as illustrated in FIG.~\ref{figure-flow}. The data flows through these components sequentially. The vectorized geometry of the input structures is first processed by the embedding module, which incorporates elemental information to enrich the feature space. Specifically, the embedding module employs two separate multilayer perceptron networks to independently embed $\Phi^{(1)}$ and $\Phi^{(2)}$ as follows:
\begin{gather}
\Phi = \text{mlp}(\Phi^{(1)}) ~\|~ \text{mlp}(\Phi^{(2)}), \label{eq:mlp_concat} \\
\text{where} \quad
\text{mlp}(x) = f_L \circ f_{L-1} \circ \cdots \circ f_1(x), \label{eq:mlp_def} \\
f_l(x) = \tanh(W_l x + b_l). \label{eq:mlp_layer}
\end{gather}
After embedding, the resulting features are either directly fed into the fitting module or passed through one or more attention modules before reaching the fitting stage.

The core of each attention module is a multi-head self-attention layer. From a physical perspective, it can be viewed as a weighted reconstruction of the local atomic neighborhood, where the model learns to infer the relative contributions of neighboring atoms. We adopt the standard formulation\cite{vaswani2017attention} for constructing the query, key, and value matrices, given by
\begin{gather}
    Q = x W^Q,\quad K = x W^K,\quad V = x W^V \\
    h = \text{softmax}(S) \cdot V
\end{gather}
However, we employ a slightly modified score function,
\begin{equation}
    S_{ij} = f_c(r_{ij}) \cdot cos(\boldsymbol{q_i}, \boldsymbol{k_j}) / \tau,
\end{equation}
where $\tau$ is a trainable scalar constrained to be greater than 0.01, introduced to automatically scale the attention scores and yield smoother weighting. 
The fitting module, composed of a multilayer perceptron, simply receives the processed features and pools them to predict atomic energies, as defined by the formulation in Eqs.~\ref{eq:mlp_def} and \ref{eq:mlp_layer}.

In summary, the ACNN architecture offers a flexible and straightforward framework for implementing and switching between two distinct MLIP designs, both of which satisfy the accuracy requirements for CSP. The primary distinction lies in the trade-off between computational cost and accuracy. 
Introducing attention modules increases model complexity and the number of trainable parameters, thereby enhancing the learning capacity of ACNN. Furthermore, stacking multiple attention modules implicitly introduces an aggregative effect analogous to that of graph neural networks (GNNs) \cite{gilmer2017neural}, which extends the effective interaction range among atoms. However, this comes at the cost of higher computational overhead and fundamentally limits the applicability of domain decomposition algorithm for efficient parallelization. Therefore, the attention-based ACNN architecture does not meet the computational requirements for large-scale, multi-component crystal structure prediction tasks of the complex systems considered in this study. Detailed accuracy and computational speed tests are provided in Section I of the Supplementary Materials.


\subsection{Automatic self-optimizing}

As a data-driven paradigm for interatomic potentials, the performance of ACNN highly depends on the quality of the sampling data.
self-optimizing aims to maintain the predictive accuracy and generalization of the ACNN model during the CSP process, while reducing the number of expensive DFT calculations. According to the principle of maximum information gain, two types of structures are particularly necessary to trigger DFT calculations and should be included in the training set. The first type consists of energy-minimum structures. These samples help reduce the discrepancy between ACNN predictions and DFT reference values, guiding the model toward a more accurate surrogate of the DFT PES. The second stems from unphysical local energy minima in ACNN, typically caused by the model’s insufficient representation of certain regions, which necessitate further refinement to avoid misleading the global optimization process in CSP.

The flowchart of the self-optimizing pipeline is illustrated in FIG.~\ref{figure-main}, where two types of data acquisition discussed above are naturally integrated.  Upon triggering the workflow, a small set of randomly generated structures undergoes DFT self-consistent field (SCF) calculations, providing the initial training dataset for ACNN (FIG.~\ref{figure-main}a). After training, CSP approach is carried out on the coarse model until the number of unphysical structures exceeds a threshold ($B_{max}$), indicating that the model’s generalization capability is no longer sufficient for reliable exploration. The energy-minimized structures obtained from relaxation are fed into a hierarchical processing workflow (FIG.~\ref{figure-main}c), beginning with fingerprint-based deduplication\cite{wang2012calypso, Pickard_2011} and interatomic distance checks to remove configurations with unrealistically close atoms prior to DFT calculations. Based on these filtered structures, a thermodynamic phase diagram is constructed using ACNN-predicted energies. Because the ACNN retains intrinsic prediction errors and the structural search is not yet exhaustive, the resulting phase diagrams may be suboptimal or incomplete. Nevertheless, they provide valuable guidance for selecting sampling points on the PES, as low-energy structures near the convex hull typically yield the most informative data for ACNN refinement.

The self-optimizing workflow proceeds iteratively, gradually improving the accuracy and generalization of the ACNN model. As its performance advances, more structures are discovered and incorporated into the training set, further reinforcing the model’s capability, as schematically illustrated in FIG.~\ref{figure-main}b. Since the sampled dataset also serves as the target for structure prediction, the resulting phase diagram or energy sequence at DFT-level accuracy can be directly constructed from the training data. After sufficient iterations, the phase diagram tends to converge, indicating that the search is approaching exhaustion. Once no additional high value structures are identified, the workflow can be terminated.

Since prediction errors are intrinsic to MLIPs, evaluating their influence on the reliability of structure prediction is essential. From the perspective of uncertainty quantification, the error can be decomposed into epistemic and aleatoric components\cite{rocken2024accurate}. By systematically sampling underexplored regions of PES, the self-optimizing scheme provides an effective mechanism for reducing epistemic uncertainty, the associated errors can be quantitatively assessed on test sets composed of labeled structures from subsequent iterations. Sources of aleatoric uncertainty mainly contributed by the intrinsic training error of the ACNN approach. Efforts to reduce this uncertainty involve preferentially sampling high-value regions, such as local minima, while avoiding configurations that are far from force equilibrium.
Collectively, these errors ultimately manifest as inaccuracies in the predicted energies, forces, and stresses during structure optimization. The energy error, $\varepsilon_E$, propagates into the calculation of formation enthalpies, thereby affecting the relative stability of structures within the phase diagram. The force error, $\varepsilon_F$, determines the equilibrium configurations obtained via structural optimization and influences the extent of structural distortion. The stress error, $\varepsilon_W$, manifests in deviations of lattice parameters and cell geometry, and further impacts the enthalpy through its contribution to the $PV$ term. If these errors are viewed as perturbations to the equilibrium structures, the resulting phase diagrams tend to overestimate the enthalpy of the structures. To achieve more accurate phase boundaries, it is often necessary to perform full DFT structural relaxations for the structures near the convex hull.


\subsection{Stoichiometric filtering}
Another major obstacle in assessing the thermodynamic stability of structures in multi-component systems is the combinatorial explosion of possible stoichiometries, which grows rapidly with the number of constituent elements. For instance, limiting the unit cell to a maximum of 30 atoms results in 269 unique compositions in binary systems. This number increases to 3,196 for ternaries and surges to 22,619 for quaternaries, rendering exhaustive exploration of all compositions virtually impossible. Under moderate pressure, the compositional space is often constrained by chemical priors such as valence rules or charge neutrality, or simplified by reducing high-dimensional phase diagrams to pseudobinary joints\cite{liu2021copex}. However, such assumptions may no longer hold under extreme conditions, where unconventional bonding and non-stoichiometric phases can emerge. One notable example is the formation of electride compounds, in which excess electrons localize in interstitial sites rather than being associated with specific atoms. 

For systems whose stoichiometries in the vast compositional space cannot be simplified a priori, we introduce a sampling strategy toward predicted low enthalpy regions near the convex hull to prioritize compositions more likely to yield stable structures for more exhaustive exploration.
Sepcificlly, after ACNN-based relaxation of a large number of structures, a convex hull analysis is performed. The sampling range is defined as the top $C_{\text{max}}$ compositions with the lowest convex hull enthalpies ($E_{\text{hull}}$), each contributing up to $L_{\text{max}}$ of their lowest enthalpy structures. These selected structures, along with those that failed to converge during relaxation, are subsequently subjected to DFT calculations. This flexible strategy introduces no prior assumptions and can therefore adapt dynamically to different systems. Moreover, the sampling bias only influences the allocation of DFT refinements across compositions without biasing structural exploration, thus that its impact on the search process remains limited and controlled.

\begin{figure*}[t]
    \includegraphics[width=2\columnwidth]{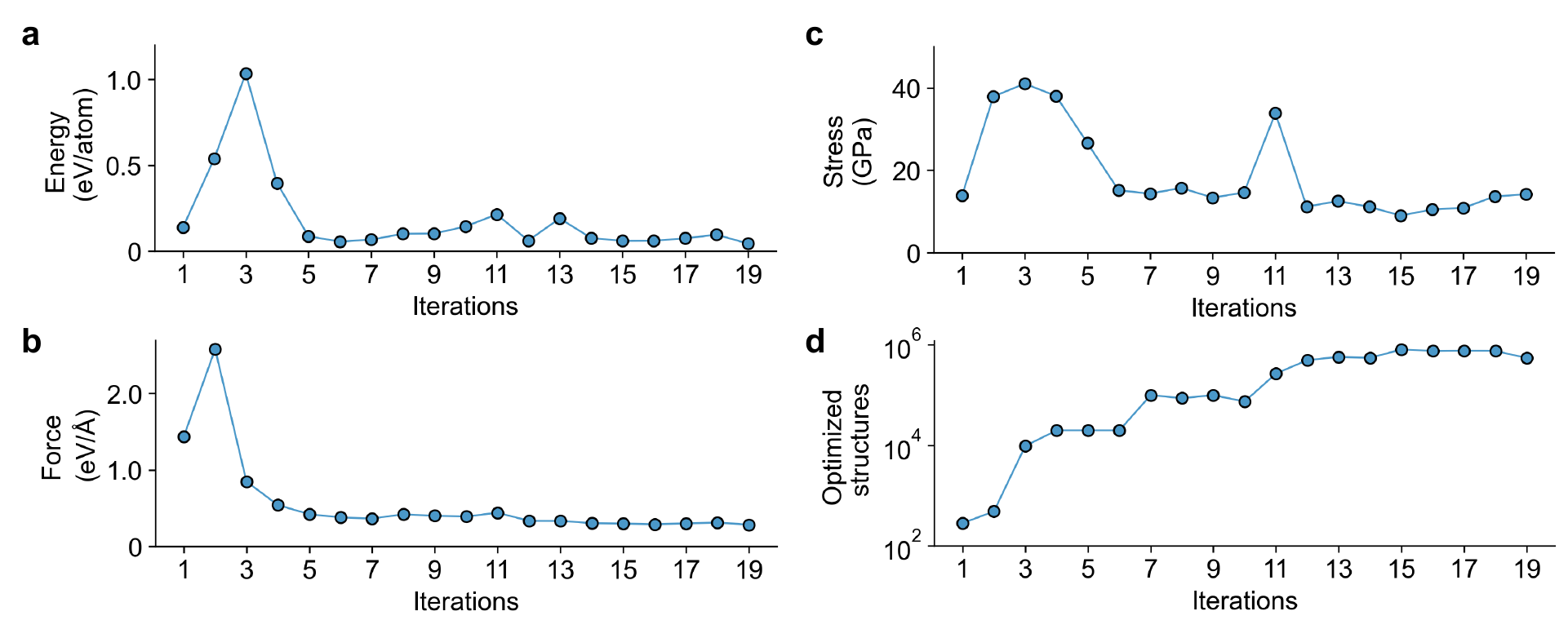}
    \centering
    \caption{ \textbf{Performance of ACNN on the Mg–Ca–H testing system}
    \textbf{a-c} Root mean square errors (RMSEs) for energy, force, and stress predictions at each iteration of the ACNN model. The errors are evaluated using DFT results from the next generation of sampled structures and tested against the model from the current iteration. \textbf{d} Number of optimized structures obtained in each iteration.
    }
    \label{figure-mgcah1}
\end{figure*}

\subsection{\label{MgCaH}Mg-Ca-H system}



\begin{figure*}[t]
    \includegraphics[width=2\columnwidth]{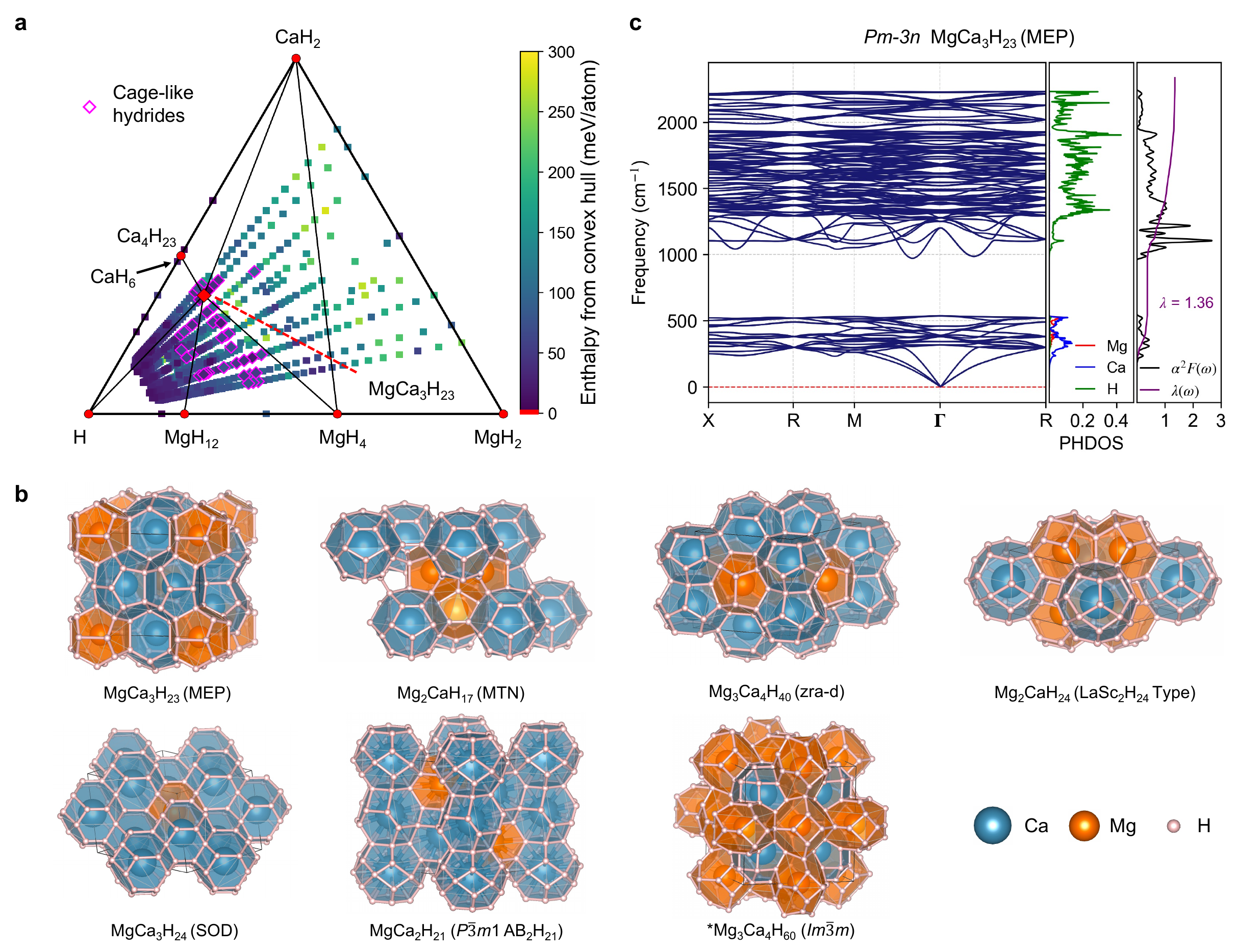}
    \centering
    \caption{ 
    \textbf{Summary of the predicted cage-like hydrides.}
    \textbf{a}  Pseudo-ternary phase diagram of MgH$_2$–CaH$_2$–H, depicting all structures with enthalpy above the convex hull below 300 meV/atom. Cage-like indicated are indicated by magenta diamonds.
    \textbf{b} Prototype structures of the predicted cage-like hydrides. Some correspond to known frameworks from zeolite databases and are labeled accordingly after their chemical formulas. Previously unreported structures are marked with an asterisk ($*$).
    \textbf{c}  Phonon dispersion, phonon density of states (PHDOS) projected onto individual atomic species, Eliashberg spectral function $\alpha^2F(\omega)$, and electron–phonon coupling strength $\lambda(\omega)$ for the thermodynamically stable $Pm\bar{3}n$ MgCa$_3$H$_{23}$. Phonon calculations show no imaginary frequencies, suggesting dynamical stability.
    }
    \label{figure-mgcah2}
\end{figure*}

We first investigated the capability of the ACNN-based CSP approach by focusing on a variable ternary system.
This choice is motivated by recent  breakthroughs in superconductivity research, where clathrate-type hydride compounds such as $\text{CaH}_6$\cite{wang2012superconductive, ma2022high} and $\text{LaH}_{10}$\cite{peng2017hydrogen, liu2017potential, drozdov2019superconductivity} have been theoretically predicted under high pressure and subsequently confirmed through experimental synthesis. These materials, known as ``metal superhydrides'', feature a distinctive structural motif where metal atoms are encapsulated within hydrogen-based clathrate cages. As one of the pioneering examples, $\text{CaH}_6$ exhibits a high superconducting critical temperature ($T_c$) of 215 K at 172 GPa. Theoretical frameworks suggest that lighter metal atoms exhibit lower frequency vibrational modes that favor superconductivity, along with higher electron-phonon coupling from weaker multicentered hydrogen bonding interactions\cite{snider2021synthesis}. These insights have motivated the incorporation of Mg into the Ca–H binary system, forming the ternary Mg–Ca–H system as a potential platform for high-temperature superconductivity.


\begin{figure*}[t]
    \includegraphics[width=2\columnwidth]{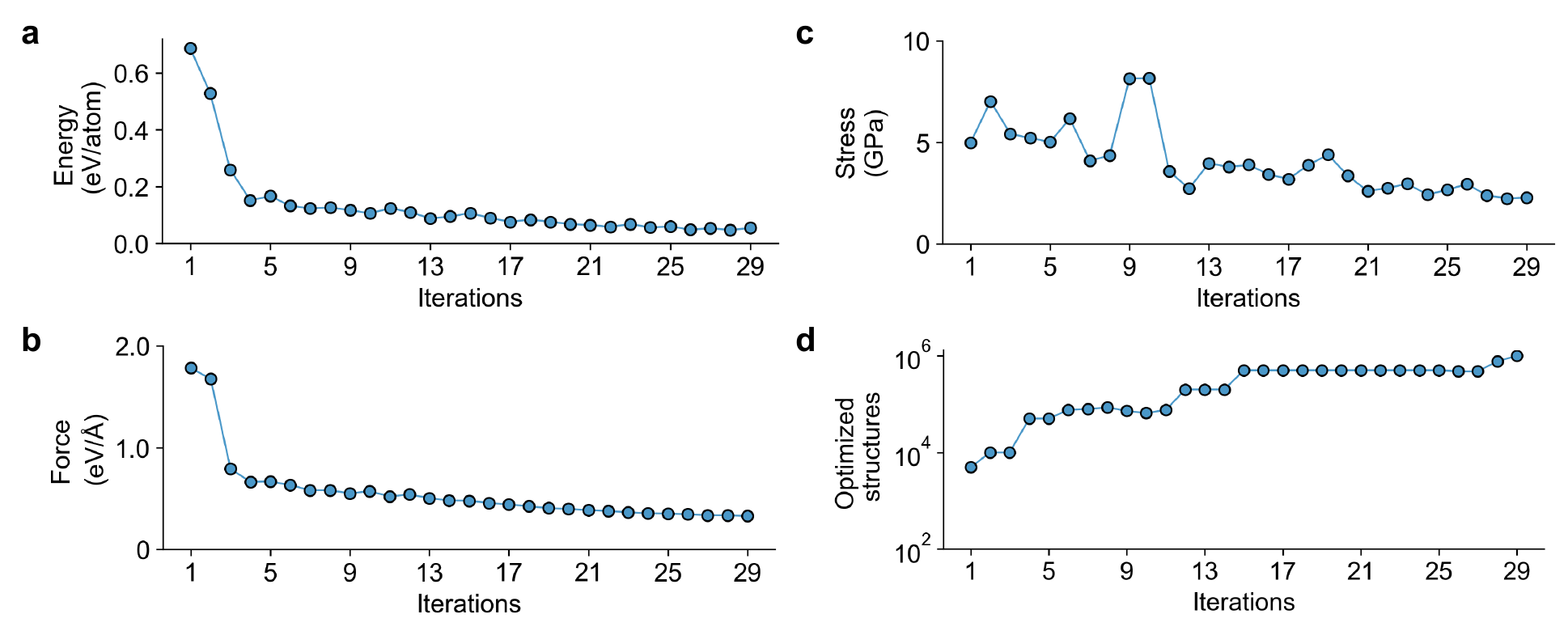}
    \centering
    \caption{ \textbf{Performance of ACNN on the quaternary Be–P–N–O system}
    \textbf{a-c} Root mean square errors (RMSEs) for energy, force, and stress predictions at each iteration of the ACNN model. The errors are evaluated using DFT results from the next generation of sampled structures and tested against the model from the current iteration. \textbf{d} Number of optimized structures obtained in each iteration.
    }
    
    \label{figure-bpno1}
\end{figure*}

Previous study by \textcite{sukmas2020near} reported a metal-substituted phase, Mg$_{0.5}$Ca$_{0.5}$H$_6$, derived from the novel $Im\bar{3}m$ CaH$_6$ prototype. This phase was predicted to exhibit near room-temperature superconductivity, with an estimated $T_c$ of 288 K at 200 GPa. More recently, \textcite{cai2023superconductivity} successfully synthesized Ca-Mg-based ternary superhydrides using binary intermetallic CaMg$_2$ and 1:1 Ca-Mg mixture as starting reactants. The resulting compounds exhibited $T_c$ approaching 168 K at 310 GPa in the CaMg$_2$-based superhydride and 182 K in 1:1 the Ca-Mg superhydride at 324 GPa. These findings indicate the existence of high $T_c$ compound in this system. However, due to the prohibitive computational cost of exploring the vast ternary space, a comprehensive ternary phase diagram remains elusive .

Herein, we conduct a structure search at 300 GPa. For the composition space, we restrict the number of Ca and Mg atoms to fewer than 4 each, with the maximum hydrogen-to-metal atomic ratio capped at 16. This resulting in a total of 827 irreducible compositions. Additionally, a constraint of a maximum of 80 atoms per unit cell was imposed, primarily to ensure the inclusion of complex cage-type hydride structures. These settings ensure a sufficiently broad search space, guaranteeing the target structure falls within the search range. Finally, leveraging the high computational efficiency of the ACNN, a total of 5,964,338 structures were optimized during the CSP stage within approximately 2.7 days on a 192-core AMD EPYC 9654 CPU node, achieving a throughput of 474 structures per core-hour.

Following the self-optimizing protocol, DFT SCF simulations were performed on an initial set of 500 randomly generated structures to initiate the computational workflow. The workflow was iteratively executed for a total of 19 cycles. In each iteration, 300 compositions exhibiting the lowest predicted $E_{\text{hull}}$ values ($C_\text{max} = 300$) were first selected. From each of these compositions, the 3 lowest energy structures ($L_\text{max} = 3$) were then chosen as PES sampling candidates for subsequently DFT calculations. In other words, this regulates the computationally expensive yet necessary DFT calculation to a maximum of 900 optimal structures per iteration. The prediction errors for energy, force, and stress, evaluated by comparing the dataset from the $\mathbf{n}$-th iteration with the ACNN model trained in the $\mathbf{(n-1)}$-th iteration, are shown in FIG.~\ref{figure-mgcah1}a-c. This validation scheme quantitatively reflects the performance of the ACNN model in predicting potentially unseen local minima structures. The overall decreasing trend in validation error indicates the model’s improving generalization capability with each iteration. By the final iteration, the ACNN achieved a root mean square error (RMSE) of 44 meV/atom for energy, 283 meV/\AA\ for force, and 14 GPa for cell stresses predictions. This continuous improvement is further evidenced by the progressive increase in the number of uninterrupted structure optimization sequences (as shown in FIG.~\ref{figure-mgcah1}d).

\begin{figure*}[t]
    \includegraphics[width=2\columnwidth]{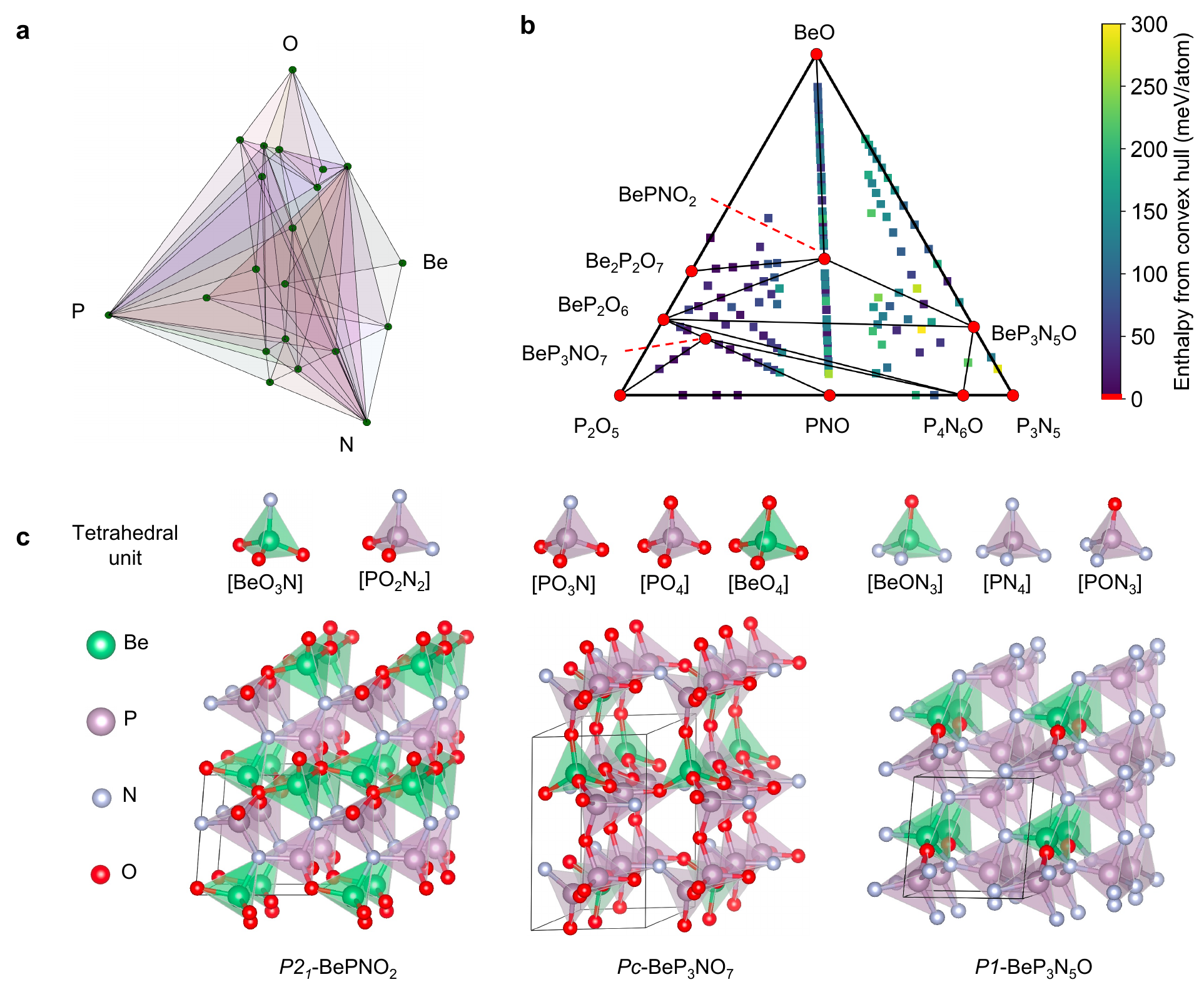}
    \centering
    \caption{ \textbf{Phase diagram and structure of Be–P–N–O system.}
    \textbf{a}  A phase diagram incorporating known structures for the quaternary system.
    \textbf{b}  Pseudo-ternary phase diagram of the BeO–P$_2$O$_5$–P$_3$N$_5$, in which three thermodynamically stable phases $P2_1$ BePNO$_2$, $Pc$ BeP$_3$NO$_7$, and $P1$ BeP$_3$N$_5$O were located. 
    \textbf{c}  Three-dimensional (3D) frameworks of thermodynamically stable quaternary Be–P–N–O phases constructed from PN$_x$O$_{4-x}$ and BeN$_x$O$_{4-x}$ (x = 1, 2, 3, 4) tetrahedral units.
    }
    \label{figure-bpno2}
\end{figure*}

Finally, 16, 901 structures were automatically sampled for DFT calculations, including 1, 115 structures necessary to correct unphysical descriptions of the ACNN model,  and the remaining 15, 786 nearly local minima structures identified as high value candidates. Among these, 8, 994 structures exhibited energy above the convex hull ($E_{hull}$) below 100 meV/atom. For convex hull refinement, we performed DFT structural relaxation on the 1,000 most stable configurations, with a maximum $E_{hull}$ of approximately 49 meV/atom. As a result, we obtain the phase diagram shown in FIG.~\ref{figure-mgcah2}a, where a thermodynamically stable $Pm\bar{3}n$ MgCa$_3$H$_{23}$ is successfully located. This structure is also dynamically stable, as confirmed by the phonon spectrum in FIG.~\ref{figure-mgcah2}c. It features a hydrogen framework adopting the `MEP' zeolite-type topology \cite{Baerlocher2017}, with Mg and Ca occupying the 2a and 6d Wyckoff sites, respectively. Alternatively, it can also be viewed as a Mg-substituted variant of the recently reported Ca$_4$H$_{23}$ phase\cite{an2024type}, suggesting its a potential high-temperature superconductor. Here, we found the estimated $T_c$ of 137 K at 300 GPa.


Due to the characteristic features of hydride structures, ideal metal superhydride crystal phases are often accompanied by numerous hydrogen-deficient structures, which typically exhibit comparable formation enthalpies.
To efficiently expand the pool of candidate clathrate hydrides, we adopt the screening strategy developed by \textcite{jiang2025data}. This strategy employs an automated topological analysis to identify and classify prototype hydrogen cage motifs. Interestingly, the result accommodates multiple metastable structural prototypes previously reported in zeolite database\cite{baerlocher2017database, rcsr_database} or other metal hydride systems\cite{he2024predicted} (see FIG.~\ref{figure-mgcah2}b). These structures and their corresponding defect configurations are marked by magenta hollow diamonds in the phase diagram shown in FIG.~\ref{figure-mgcah2}a.
We predict a new metastable $Im\bar{3}m$ Mg$_3$Ca$_4$H$_{60}$ phase with an $E_{hull}$ of 66 meV/atom at 300 GPa. In this structure, Mg and Ca atoms occupy the 6b and 8c Wyckoff sites respectively, forming two unusual Mg-centered $H_{24}$ cage with $D_{4v}$ point group symmetry and Ca-centered $H_{30}$ cage with $D_{3d}$ symmetry. The H–H bond lengths within these cages measure 0.94, 1.10, 1.11, 1.17 and 1.19 \AA\ at 300 GPa, comparable to those observed in known superconducting hydrogen-rich clathrates, such as CaH$_6$ (1.24 \AA\ at 150GPa)\cite{wang2012superconductive} and LaH$_{10}$ (1.07 and 1.16 \AA\ at 300 GPa)\cite{peng2017hydrogen, liu2017potential}. However, phonon calculations indicate that the structure is dynamically unstable.


\begin{table}[ht]
    \centering
    \caption{
    \textbf{Superconducting properties of dynamical stable hydrides.} 
    For each hydrogen cage prototype, the most energetically favorable defect-free structure is presented.
    Listed are the electron–phonon coupling strength $\lambda$, the logarithmic average phonon frequency $\omega_\mathrm{log}$, and the estimated superconducting transition temperature $T_c$.
    }
    \renewcommand{\arraystretch}{1.3}
    \begin{tabular}{lccccc}
        \toprule
        \textbf{Prototype} & \textbf{Hydride} & $\boldsymbol{E_\mathrm{hull}}$ (meV/atom) & $\boldsymbol{\lambda}$ & $\boldsymbol{\omega_\mathrm{log}}$ (K) & $\boldsymbol{T_c}$ (K) \\
        \midrule
        MEP & MgCa$_3$H$_{23}$ & 0 & 1.36 & 1311 & 137 \\
        SOD & MgCa$_3$H$_{24}$ & 39 & 0.81 & 1810 & 86 \\
        SOD & Mg$_2$CaH$_{18}$ & 51 & 1.44 & 1595 & 175 \\
        $P\bar{3}m1$ AB$_2$H$_{21}$ & MgCa$_2$H$_{21}$ & 43 & 2.05 & 1365 & 200 \\
        \bottomrule
    \end{tabular}
    \label{tab:Tc}
\end{table}

Building upon the identified candidate structures, we further estimated their phonon properties (Section III of the Supplementary Materials) and superconducting critical temperatures ($T_c$) using the Allen-Dynes modified McMillan formula. The estimated $T_c$ values for the dynamically stable hydrides are presented in TABLE~\ref{tab:Tc}, revealing numerous promising high-temperature superconductors that may be experimentally accessible.  These findings underscore the effectiveness of our algorithm in discovering complex superconducting hydrides with favorable properties.

\begin{figure*}[t]
    \includegraphics[width=2\columnwidth]{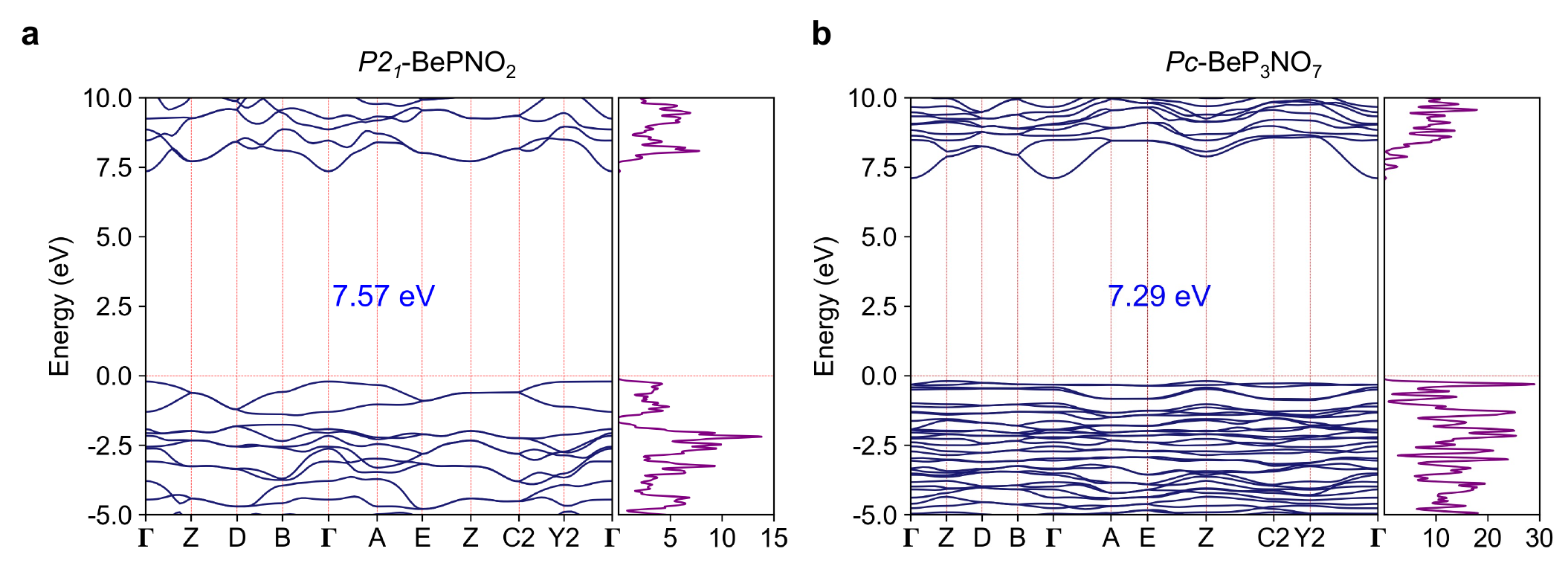}
    \centering
    \caption{ \textbf{HSE-calculated electronic band structures and density of states (DOS).}
    \textbf{a} $P2_1$ BePNO$_2$ features a direct band gap of 7.57 eV (164 nm).
    \textbf{b} $Pc$ BeP$_3$NO$_7$ exhibits an indirect band gap of 7.29 eV (170 nm).
    }
    \label{figure-bpno3}
\end{figure*}

\subsection{\label{BePNO}Be-P-N-O system}

In the second test, we apply our method to a more challenging CSP task by focusing on quaternary compounds.  Quaternary system typically present a significantly more complex structural search space due to their greater compositional and configurational degrees of freedom. A particularly promising application lies in the discovery of nonlinear optical (NLO) crystals. As a classical system, phosphates exhibit structural diversity, making them strong candidates for the discovery of new nonlinear optical (NLO) materials \cite{yu2017m4mg4, li2016new} capable of generating coherent UV/DUV light through laser frequency conversion. These materials are crucial for applications such as ultra-high-resolution lithography, photochemical synthesis, and high-precision micromachining.
Unfortunately, most of them suffer from low SHG coefficients and poor birefringence due to the small microscopic hyperpolarizability and polarizability anisotropy of the nonpolar tetrahedral [PO$_4$] group. In particular, the small birefringence hampers the phase-matching (PM) ability in the UV/DUV region and is also the key problem in most traditional alkali/alkaline earth metal phosphate materials. [PN$_x$O$_{4-x}$] (x = 1, 2, 3) groups, formed by introducing N element into the [PO$_4$] group, possessing a balanced polarizability anisotropy, hyperpolarizability and a relatively large the highest occupied molecular orbital (HOMO)-the lowest unoccupied molecular orbital (LUMO) gap, have been proved as good functional groups for NLO crystals and applied in PNO crystal successfully\cite{xie2022pno}. Herein, we integrate this new NLO active group [PN$_x$O$_{4-x}$] (x = 1, 2, 3) with alkaline earth metal, taking the Be-P-N-O system as an example, to systematically predict crystal structures using our method. Given the advantages of large band gap for Be element and balanced micro properties for [PN$_x$O$_{4-x}$] (x = 1, 2, 3) groups, the Be-P-N-O system is deemed a suitable test case for exploring NLO materials and verifying our method.


Following the same procedure used for the above ternary system, we first established the compositional search space. A key distinction, however, is that rigid charge balance rules under ambient pressure can effectively constrain the compositional search space. To ensure both comprehensive and efficient exploration while incorporating experimentally feasible synthesis pathways, we first systematically constructed pseudo-binary phase diagrams based on known binary and ternary compounds, from which all appearing compositions were incorporated into the CSP search space. To further enrich this space, we added all charge neutral compositions with fewer than 32 atoms per unit cell. In total, 373 irreducible compositions were identified and subsequently scaled to multiple formula units, with unit cell sizes constrained to fewer than 60 atoms.

A total of 9,381,987 structures were fully relaxed during the CSP stage by ACNN model, executed over 29 self-optimizing iterations to efficiently explore the complex configurational landscape. The entire relaxation process took approximately 1.8 days on a 192-core AMD 9654 CPU, achieving a speed of about 1131 structures per core-hour. Improved relaxation efficiency relative to the previous ternary system primarily attributed to the reduced average number of atoms per structure, as well as the ambient pressure conditions and larger atomic radii (relative to hydrogen) decrease the number of neighboring atoms, collectively lowering the computational workload. For structure sampling, a more flexible strategy was employed. During the initial five iterations, two distinct structures with the lowest energy ($L_\text{max} = 2$) were selected from each of the 270 compositions ($C_\text{max} = 270$) with the lowest predicted $E_\text{hull}$, aiming to facilitate broader exploration of the compositional space. In subsequent iterations, the sampling parameters were adjusted to $L_\text{max} = 4$ and $C_\text{max} = 95$ to more densely probe novel structures within the compositional region that had been largely validated as near the convex hull and potentially thermodynamical stable.

The corresponding prediction errors during each iterations are presented in FIG.~\ref{figure-bpno1}a-d. The quaternary system exhibits a more gradual decline in validation error, consistent with the observed trend reported by \textcite{wang2024concurrent} when transitioning from binary to ternary systems. This slower convergence likely arises from the increased structural complexity of higher-order systems, which imposes greater demands on both CSP and the generalization capacity of the MLIP. The ACNN model ultimately achieves a RMSE of 62 meV/atom for energy, 325 meV/\AA\ for forces, and 2.2 GPa for cell stresses predictions.

A total of 13,214 structures were selected for SCF calculations, comprising 11,460 local minima and 1,754 unphysical configurations mistakenly predicted as low-enthalpy by ACNN. Among these, top 1,000 structures with the lowest $E_{hull}$ values were subjected to full DFT structural optimization to construct a more accurate convex hull surface. The complete quaternary phase diagram is presented in the FIG.~\ref{figure-bpno2}a. Three thermodynamically stable phases were identified: $P2_1$-BePNO$_2$, $Pc$-BeP$_3$NO$_7$, and $P_1$-BeP$_3$N$_5$O. These phases are distributed within the pseudo-ternary compositional space spanned by BeO-P$_2$O$_5$-P$_3$N$_5$, as shown in FIG.~\ref{figure-bpno2}b. The crystal structures of all three phases are composed of P-centered and Be-centered tetrahedra, interconnected through corner sharing to form robust three-dimensional (3D) frameworks with distinctive cavity geometries (see Fig.~\ref{figure-bpno2}c). This structure structural motif resembles those reported in prior studies, where nitrogen atoms partially substitute oxygen within phosphate frameworks, forming PN$_x$O$_{4-x}$ (x = 1, 2, 3, 4) mixed anionic tetrahedra. Such hybrid coordination environments contribute to the structural diversity and potential functionality of these materials.

The electronic band structures of the thermodynamically stable are presented in FIG.~\ref{figure-bpno3}a,b, revealing band gaps of 7.57 eV (164 nm) and 7.29 eV (170 nm), respectively. Their shortest PM wavelengths were calculated to be 260 nm and 334 nm, indicating potential for ultraviolet optical applications.  Additionally, the resulting structure pool include nearly 3,000 structures with formation enthalpies within $E_{hull} < 100 ~\text{meV/atom}$, commonly associated with potential experimental synthesizability. A more comprehensive analysis of the physical properties of these metastable candidates will provided in a separate forthcoming work.


\section{Discussion}

We introduce ACNN integrated with CSP to establish an efficient and self-optimizing workflow that addresses the high computational complexity of predicting multi-component systems. The robustness of this self-optimizing scheme, together with the effectiveness of ACNN, enables the exploration of millions of candidate structures across diverse stoichiometries and large atomic configurations, allowing a far more comprehensive investigation of the material space. We first apply the algorithm to the high pressure Mg–Ca–H system as a test case, where we discover a stable $Pm\bar{3}n$ $\text{MgCa}_3\text{H}_{23}$ phase along with a series of metastable candidates as promising high-$T_c$ superconductors. We further demonstrate its applicability to the more complex quaternary Be–P–N–O system, where two potential deep-UV nonlinear optical crystals, $P2_1$-BePNO$_2$ and $Pc$-BeP$_3$NO$_7$, are discovered. Together, these case studies corroborate the diversity of promising compounds in multi-component systems and highlight the capability of our algorithm to efficiently uncover functional materials within vast, high-dimensional chemical spaces.

Nevertheless, designing complex materials remains a longstanding scientific challenge. In practice, system specific considerations often arise. For instance, in the Be–P–N–O system, when targeting high performance nonlinear optical properties, the requirement of non-centrosymmetry may call for adjustments to the symmetry constraints used in random structure generation, whereas a more rigorous assessment of thermodynamic stability demands a broader sampling of the structural space. Similarly, when searching very large crystal structures, biasing the initial population with local ordering can provide favorable pathways toward the ordered ground state\cite{lyakhov2010predict}. To address such diverse scenarios, our self-optimizing framework is built with substantial flexibility and modularity, enabling adaptation to different applications and seamless integration with state-of-the-art structure design algorithms developed by the community.
In particular, with recent advances in AI-assisted materials design, new methods such as generative models\cite{antunes2024crystal, luo2024deep, xie2021crystal} can be readily combined with our framework to achieve more efficient structure design from the initial generation stage, or with compositional optimization approaches such as PhaseBO\cite{vasylenko2024inferring} to enable more effective sampling by reducing the dimensionality of the compositional space.

The ACNN software package is released in conjunction with the first principles simulation code ARES\cite{xu2019ab}, together forming a technically integrated and robust computational ecosystem. This integration fully unleashes the potential of ACNN for materials design and downstream property simulations, facilitating an closed-loop approach to materials design.
Such an integrated software design principles establish a solid computational foundation for automated materials discovery.

\section{Methods}

\subsection{Training and refining protocol}
All parameters $\boldsymbol{c}_\nu$ in Eq.~\ref{eq:site_ene} are trainable through the backpropagation process. Considering that the training set may contain structures with varying numbers of atoms, we define the loss function as follows,
\begin{equation}
    L = \frac{1}{N_B} \sum_{i=1}^{N_B} p_e \frac{|E_i - \hat{E_i}| ^2}{N_i} + p_f \frac{||F_i - \hat{F_i}|| ^2}{3 N_i} + 
        p_w \frac{||W_i - \hat{W_i}|| ^2}{3 N_i}.
\end{equation}
Here, $N_B$ is the minibatch size, and $N_i$ the number of atoms in structure $i$. Hatted quantities (e.g., $\hat{E}_i$) represent reference values, whereas unhatted ones correspond to ACNN predictions. $p_e$, $p_f$, and $p_w$ are tunable parameters that flexibly adjust the relative contributions of each component in the loss function. We employed the AdamW\cite{loshchilov2017decoupled} optimizer, which incorporates L2 regularization to mitigate overfitting by penalizing large parameter values. This may help improve the generalization performance of the ACNN model when trained on a diverse set of structures broadly sampled from the PES.

When applied to CSP, the training set typically needs to be continuously expanded through the self-optimizing process discussed below. ACNN supports restarting training or refining the model from any existing checkpoint with a new or updated dataset. During refinement, small perturbations are introduced to the trainable parameters, which helps mitigate systematic errors in extrapolated predictions. The rationale is that the perturbed model can quickly stabilize in well-sampled regions at the onset of refinement, while maintaining controlled stochasticity in out-of-distribution regions.

The ACNN hyperparameter configurations for both systems share the following settings: elemental embedding networks with [64, 64] configuration (two layers of 64 neurons), fitting networks with [256, 256] configuration (two layers of 256 neurons), radial basis functions of order $n_{max}^r = 16$, angular basis functions of order $n_{max}^a = 14$, and an angular momentum cutoff of $l_{max} = 4$ for characterizing single-bond interactions. The cutoff distances differ between systems: for the Mg-Ca-H system under high pressure, $r_c^r = 5.0 $ \AA\ and $r_c^a = 4.0 $ \AA\ are used, whereas for the Be-P-N-O system, $r_c^r = 6.5$ \AA\ and $r_c^a = 6.0 $ \AA\ .
In each iteration of the self-optimizing workflow, the ACNN is trained for $10^5$ backpropagation steps. During training, an exponential decay learning rate schedule is adopted, where the learning rate decays by a factor of 0.97 every $10^4$ steps. The loss prefactors $p_e$, $p_f$, and $p_w$ were initially set to 0.1, 10 and 0.1, respectively. While $p_e$ and $p_w$ were kept constant, $p_f$ was exponentially decayed to 0.1 using the same decay factor of 0.97 every $10^4$ steps.

\subsection{Structure generation}
Structure generation was performed using the Crystal Structure Analysis by Particle Swarm Optimization (CALYPSO) code\cite{wang2010crystal, wang2012calypso} in a purely random sampling mode, applying only stochastic symmetry constraints. This approach enables an unbiased assessment of our method by deliberately excluding the global structural evolution operators commonly integrated into conventional CSP workflows, which, although often statistically efficient, can introduce method-dependent biases.

Minimum interatomic distances and cell volumes are specified to prevent the generation of unphysical configurations at the outset. For the Mg–Ca–H system, the initial minimum distances (\AA) are given by the symmetric matrix,
\begin{equation}
    d_{\mathrm{Mg\text{-}Ca\text{-}H}} =
\begin{pmatrix}
\text{Mg} & \text{Ca} & \text{H} \\
1.9 & 1.9 & 1.4 \\
& 1.9 & 1.6 \\
&     & 0.8
\end{pmatrix},
\end{equation}
where the matrix is symmetric with rows and columns ordered as (Mg, Ca, H). The initial cell volume is estimated as a linear combination of atomic contributions,
\begin{equation}
    (V_\mathrm{Mg}, V_\mathrm{Ca}, V_\mathrm{H}) = (7.1,\ 8.8,\ 1.4) \; \mathrm{\AA}^3
\end{equation}
For the Be–P–N–O system, the initial minimum distances (Å) are given by,
\begin{equation}
    d_{\mathrm{Be\text{-}P\text{-}N\text{-}O}} =
\begin{pmatrix}
\text{Be} & \text{P} & \text{N} & \text{O} \\
  1.9  & 1.9  & 1.4  & 1.3 \\
& 1.6  & 1.4  & 1.5 \\
&      & 0.9  & 0.9 \\
&      &      & 1.0
\end{pmatrix},
\end{equation}
with rows and columns ordered as (Be, P, N, O). The atomic volumes are 
\begin{equation}
    (V_\mathrm{Be}, V_\mathrm{P}, V_\mathrm{N}, V_\mathrm{O}) = (7.9,\ 27.3,\ 25.8,\ 14.5) \; \mathrm{\AA}^3.
\end{equation}

\subsection{Calculation details}
All DFT calculations were performed using the Ab initio atomic mateRial modEling Software (ARES) package \cite{xu2019ab}, within the framework of the generalized gradient approximation level of theory\cite{perdew1996generalized}, and employing the optimized norm-conserving Vanderbilt pseudopotentials \cite{hamann2013optimized, van2018pseudodojo} to account for core-electron effects.
For the Mg-Ca-H system, the plane-wave and density cutoff energies were set to 70 Ry and 280 Ry, respectively, while for the Be-P-N-O system, they were set to 80 Ry and 320 Ry. Both systems share the same k-point spacings configuration, with $2\pi \times 0.03 A^{-1}$ for SCF calculations during structure prediction and $2\pi \times 0.02 A^{-1}$ for electronic structure calculations.

To achieve optimal computational efficiency of ACNN, we employ an in-house implementation of the Broyden–Fletcher–Goldfarb–Shanno (BFGS) algorithm \cite{nocedal2006numerical} to perform batch structure optimization. 
Atomic positions and lattice parameters are optimized using Hellmann–Feynman forces and the computed stress tensor, respectively, both derived from the ACNN potential. Convergence was achieved when the enthalpy change fell below $ 10^{-6}$ eV/atom and the maximum atomic force was less than $10^{-3}$ eV/\AA\ . Here, the batch mode enables multiple structures to be optimized simultaneously, significantly increasing the degree of parallelism.

\section*{Competing Interests}
The authors declare no conflict of interest.

\section*{Data Availability}
Some data supporting the findings of this study are available in the Supplementary Materials. Additional data are available from the corresponding author upon reasonable request.

\section*{Code Availability}
The custom code developed and used in this study will be made publicly available upon publication. Prior to that, it can be obtained from the corresponding author upon reasonable request.

\section*{Author contributions}
Y.X., H.L., C.X., and J.L. designed the research.
J.L., J.L., and X.Z. developed the code.
J.L. conducted the majority of the CSP calculations.
J.L., Q.S., J.F., and B.J. performed the property calculations.
J.L., J.L., H.L., K.B., C.X., Y.X., and Y.M. analyzed and interpreted the data and contributed to writing the manuscript.

\begin{acknowledgments}
This work is supported by the National Key R\&D Program of China (Grant No. 2023YFB3003001), the National Natural Science Foundation of China (Grant No. 12374008 and 12022408), the National Natural Science Foundation of China (52403305), the Strategic Priority Research Program of the Chinese Academy of Sciences (XDB0880000), the Interdisciplinary Integration and Innovation Project of JLU, the Fundamental Research Funds for the Central Universities and the Program for JLU Science and Technology Innovative Research Team (JLUSTIRT). We also thank the Changchun Computing Center and the Eco-Innovation Center for providing comprehensive computing resources and technical support throughout the completion of this work.
\end{acknowledgments}


\bibliography{apssamp}

\begin{thebibliography}{67}%
\makeatletter
\providecommand \@ifxundefined [1]{%
 \@ifx{#1\undefined}
}%
\providecommand \@ifnum [1]{%
 \ifnum #1\expandafter \@firstoftwo
 \else \expandafter \@secondoftwo
 \fi
}%
\providecommand \@ifx [1]{%
 \ifx #1\expandafter \@firstoftwo
 \else \expandafter \@secondoftwo
 \fi
}%
\providecommand \natexlab [1]{#1}%
\providecommand \enquote  [1]{``#1''}%
\providecommand \bibnamefont  [1]{#1}%
\providecommand \bibfnamefont [1]{#1}%
\providecommand \citenamefont [1]{#1}%
\providecommand \href@noop [0]{\@secondoftwo}%
\providecommand \href [0]{\begingroup \@sanitize@url \@href}%
\providecommand \@href[1]{\@@startlink{#1}\@@href}%
\providecommand \@@href[1]{\endgroup#1\@@endlink}%
\providecommand \@sanitize@url [0]{\catcode `\\12\catcode `\$12\catcode `\&12\catcode `\#12\catcode `\^12\catcode `\_12\catcode `\%12\relax}%
\providecommand \@@startlink[1]{}%
\providecommand \@@endlink[0]{}%
\providecommand \url  [0]{\begingroup\@sanitize@url \@url }%
\providecommand \@url [1]{\endgroup\@href {#1}{\urlprefix }}%
\providecommand \urlprefix  [0]{URL }%
\providecommand \Eprint [0]{\href }%
\providecommand \doibase [0]{https://doi.org/}%
\providecommand \selectlanguage [0]{\@gobble}%
\providecommand \bibinfo  [0]{\@secondoftwo}%
\providecommand \bibfield  [0]{\@secondoftwo}%
\providecommand \translation [1]{[#1]}%
\providecommand \BibitemOpen [0]{}%
\providecommand \bibitemStop [0]{}%
\providecommand \bibitemNoStop [0]{.\EOS\space}%
\providecommand \EOS [0]{\spacefactor3000\relax}%
\providecommand \BibitemShut  [1]{\csname bibitem#1\endcsname}%
\let\auto@bib@innerbib\@empty
\bibitem [{\citenamefont {Ceder}\ \emph {et~al.}(1998)\citenamefont {Ceder}, \citenamefont {Chiang}, \citenamefont {Sadoway}, \citenamefont {Aydinol}, \citenamefont {Jang},\ and\ \citenamefont {Huang}}]{ceder1998identification}%
  \BibitemOpen
  \bibfield  {author} {\bibinfo {author} {\bibfnamefont {G.}~\bibnamefont {Ceder}}, \bibinfo {author} {\bibfnamefont {Y.-M.}\ \bibnamefont {Chiang}}, \bibinfo {author} {\bibfnamefont {D.}~\bibnamefont {Sadoway}}, \bibinfo {author} {\bibfnamefont {M.}~\bibnamefont {Aydinol}}, \bibinfo {author} {\bibfnamefont {Y.-I.}\ \bibnamefont {Jang}},\ and\ \bibinfo {author} {\bibfnamefont {B.}~\bibnamefont {Huang}},\ }\bibfield  {title} {\bibinfo {title} {Identification of cathode materials for lithium batteries guided by first-principles calculations},\ }\href@noop {} {\bibfield  {journal} {\bibinfo  {journal} {Nature}\ }\textbf {\bibinfo {volume} {392}},\ \bibinfo {pages} {694} (\bibinfo {year} {1998})}\BibitemShut {NoStop}%
\bibitem [{\citenamefont {Liu}\ \emph {et~al.}(2020)\citenamefont {Liu}, \citenamefont {Chen}, \citenamefont {Wang}, \citenamefont {Liu}, \citenamefont {Jiang}, \citenamefont {Zhang}, \citenamefont {Liu},\ and\ \citenamefont {Zhou}}]{liu2020two}%
  \BibitemOpen
  \bibfield  {author} {\bibinfo {author} {\bibfnamefont {C.}~\bibnamefont {Liu}}, \bibinfo {author} {\bibfnamefont {H.}~\bibnamefont {Chen}}, \bibinfo {author} {\bibfnamefont {S.}~\bibnamefont {Wang}}, \bibinfo {author} {\bibfnamefont {Q.}~\bibnamefont {Liu}}, \bibinfo {author} {\bibfnamefont {Y.-G.}\ \bibnamefont {Jiang}}, \bibinfo {author} {\bibfnamefont {D.~W.}\ \bibnamefont {Zhang}}, \bibinfo {author} {\bibfnamefont {M.}~\bibnamefont {Liu}},\ and\ \bibinfo {author} {\bibfnamefont {P.}~\bibnamefont {Zhou}},\ }\bibfield  {title} {\bibinfo {title} {Two-dimensional materials for next-generation computing technologies},\ }\href@noop {} {\bibfield  {journal} {\bibinfo  {journal} {Nature Nanotechnology}\ }\textbf {\bibinfo {volume} {15}},\ \bibinfo {pages} {545} (\bibinfo {year} {2020})}\BibitemShut {NoStop}%
\bibitem [{\citenamefont {N{\o}rskov}\ \emph {et~al.}(2009)\citenamefont {N{\o}rskov}, \citenamefont {Bligaard}, \citenamefont {Rossmeisl},\ and\ \citenamefont {Christensen}}]{norskov2009towards}%
  \BibitemOpen
  \bibfield  {author} {\bibinfo {author} {\bibfnamefont {J.~K.}\ \bibnamefont {N{\o}rskov}}, \bibinfo {author} {\bibfnamefont {T.}~\bibnamefont {Bligaard}}, \bibinfo {author} {\bibfnamefont {J.}~\bibnamefont {Rossmeisl}},\ and\ \bibinfo {author} {\bibfnamefont {C.~H.}\ \bibnamefont {Christensen}},\ }\bibfield  {title} {\bibinfo {title} {Towards the computational design of solid catalysts},\ }\href@noop {} {\bibfield  {journal} {\bibinfo  {journal} {Nature chemistry}\ }\textbf {\bibinfo {volume} {1}},\ \bibinfo {pages} {37} (\bibinfo {year} {2009})}\BibitemShut {NoStop}%
\bibitem [{\citenamefont {De~Leon}\ \emph {et~al.}(2021)\citenamefont {De~Leon}, \citenamefont {Itoh}, \citenamefont {Kim}, \citenamefont {Mehta}, \citenamefont {Northup}, \citenamefont {Paik}, \citenamefont {Palmer}, \citenamefont {Samarth}, \citenamefont {Sangtawesin},\ and\ \citenamefont {Steuerman}}]{de2021materials}%
  \BibitemOpen
  \bibfield  {author} {\bibinfo {author} {\bibfnamefont {N.~P.}\ \bibnamefont {De~Leon}}, \bibinfo {author} {\bibfnamefont {K.~M.}\ \bibnamefont {Itoh}}, \bibinfo {author} {\bibfnamefont {D.}~\bibnamefont {Kim}}, \bibinfo {author} {\bibfnamefont {K.~K.}\ \bibnamefont {Mehta}}, \bibinfo {author} {\bibfnamefont {T.~E.}\ \bibnamefont {Northup}}, \bibinfo {author} {\bibfnamefont {H.}~\bibnamefont {Paik}}, \bibinfo {author} {\bibfnamefont {B.}~\bibnamefont {Palmer}}, \bibinfo {author} {\bibfnamefont {N.}~\bibnamefont {Samarth}}, \bibinfo {author} {\bibfnamefont {S.}~\bibnamefont {Sangtawesin}},\ and\ \bibinfo {author} {\bibfnamefont {D.~W.}\ \bibnamefont {Steuerman}},\ }\bibfield  {title} {\bibinfo {title} {Materials challenges and opportunities for quantum computing hardware},\ }\href@noop {} {\bibfield  {journal} {\bibinfo  {journal} {Science}\ }\textbf {\bibinfo {volume} {372}},\ \bibinfo {pages} {eabb2823} (\bibinfo {year} {2021})}\BibitemShut {NoStop}%
\bibitem [{\citenamefont {Walsh}(2015)}]{walsh2015quest}%
  \BibitemOpen
  \bibfield  {author} {\bibinfo {author} {\bibfnamefont {A.}~\bibnamefont {Walsh}},\ }\bibfield  {title} {\bibinfo {title} {The quest for new functionality},\ }\href@noop {} {\bibfield  {journal} {\bibinfo  {journal} {Nature chemistry}\ }\textbf {\bibinfo {volume} {7}},\ \bibinfo {pages} {274} (\bibinfo {year} {2015})}\BibitemShut {NoStop}%
\bibitem [{\citenamefont {Lu}\ \emph {et~al.}(2021)\citenamefont {Lu}, \citenamefont {Tournet}, \citenamefont {Dastafkan}, \citenamefont {Liu}, \citenamefont {Ng}, \citenamefont {Karuturi}, \citenamefont {Zhao},\ and\ \citenamefont {Yin}}]{lu2021noble}%
  \BibitemOpen
  \bibfield  {author} {\bibinfo {author} {\bibfnamefont {H.}~\bibnamefont {Lu}}, \bibinfo {author} {\bibfnamefont {J.}~\bibnamefont {Tournet}}, \bibinfo {author} {\bibfnamefont {K.}~\bibnamefont {Dastafkan}}, \bibinfo {author} {\bibfnamefont {Y.}~\bibnamefont {Liu}}, \bibinfo {author} {\bibfnamefont {Y.~H.}\ \bibnamefont {Ng}}, \bibinfo {author} {\bibfnamefont {S.~K.}\ \bibnamefont {Karuturi}}, \bibinfo {author} {\bibfnamefont {C.}~\bibnamefont {Zhao}},\ and\ \bibinfo {author} {\bibfnamefont {Z.}~\bibnamefont {Yin}},\ }\bibfield  {title} {\bibinfo {title} {Noble-metal-free multicomponent nanointegration for sustainable energy conversion},\ }\href@noop {} {\bibfield  {journal} {\bibinfo  {journal} {Chemical Reviews}\ }\textbf {\bibinfo {volume} {121}},\ \bibinfo {pages} {10271} (\bibinfo {year} {2021})}\BibitemShut {NoStop}%
\bibitem [{\citenamefont {Bednorz}\ and\ \citenamefont {M{\"u}ller}(1986)}]{bednorz1986possible}%
  \BibitemOpen
  \bibfield  {author} {\bibinfo {author} {\bibfnamefont {J.~G.}\ \bibnamefont {Bednorz}}\ and\ \bibinfo {author} {\bibfnamefont {K.~A.}\ \bibnamefont {M{\"u}ller}},\ }\bibfield  {title} {\bibinfo {title} {Possible high t c superconductivity in the ba- la- cu- o system},\ }\href@noop {} {\bibfield  {journal} {\bibinfo  {journal} {Zeitschrift f{\"u}r Physik B Condensed Matter}\ }\textbf {\bibinfo {volume} {64}},\ \bibinfo {pages} {189} (\bibinfo {year} {1986})}\BibitemShut {NoStop}%
\bibitem [{\citenamefont {Kamihara}\ \emph {et~al.}(2008)\citenamefont {Kamihara}, \citenamefont {Watanabe}, \citenamefont {Hirano},\ and\ \citenamefont {Hosono}}]{kamihara2008iron}%
  \BibitemOpen
  \bibfield  {author} {\bibinfo {author} {\bibfnamefont {Y.}~\bibnamefont {Kamihara}}, \bibinfo {author} {\bibfnamefont {T.}~\bibnamefont {Watanabe}}, \bibinfo {author} {\bibfnamefont {M.}~\bibnamefont {Hirano}},\ and\ \bibinfo {author} {\bibfnamefont {H.}~\bibnamefont {Hosono}},\ }\bibfield  {title} {\bibinfo {title} {Iron-based layered superconductor la [o1-x f x] feas (x= 0.05- 0.12) with t c= 26 k},\ }\href@noop {} {\bibfield  {journal} {\bibinfo  {journal} {Journal of the American Chemical Society}\ }\textbf {\bibinfo {volume} {130}},\ \bibinfo {pages} {3296} (\bibinfo {year} {2008})}\BibitemShut {NoStop}%
\bibitem [{\citenamefont {Song}\ \emph {et~al.}(2023)\citenamefont {Song}, \citenamefont {Bi}, \citenamefont {Nakamoto}, \citenamefont {Shimizu}, \citenamefont {Liu}, \citenamefont {Zou}, \citenamefont {Liu}, \citenamefont {Wang},\ and\ \citenamefont {Ma}}]{song2023stoichiometric}%
  \BibitemOpen
  \bibfield  {author} {\bibinfo {author} {\bibfnamefont {Y.}~\bibnamefont {Song}}, \bibinfo {author} {\bibfnamefont {J.}~\bibnamefont {Bi}}, \bibinfo {author} {\bibfnamefont {Y.}~\bibnamefont {Nakamoto}}, \bibinfo {author} {\bibfnamefont {K.}~\bibnamefont {Shimizu}}, \bibinfo {author} {\bibfnamefont {H.}~\bibnamefont {Liu}}, \bibinfo {author} {\bibfnamefont {B.}~\bibnamefont {Zou}}, \bibinfo {author} {\bibfnamefont {G.}~\bibnamefont {Liu}}, \bibinfo {author} {\bibfnamefont {H.}~\bibnamefont {Wang}},\ and\ \bibinfo {author} {\bibfnamefont {Y.}~\bibnamefont {Ma}},\ }\bibfield  {title} {\bibinfo {title} {Stoichiometric ternary superhydride labeh 8 as a new template for high-temperature superconductivity at 110 k under 80 gpa},\ }\href@noop {} {\bibfield  {journal} {\bibinfo  {journal} {Physical Review Letters}\ }\textbf {\bibinfo {volume} {130}},\ \bibinfo {pages} {266001} (\bibinfo {year} {2023})}\BibitemShut {NoStop}%
\bibitem [{\citenamefont {Otto}\ \emph {et~al.}(2013)\citenamefont {Otto}, \citenamefont {Dlouh{\`y}}, \citenamefont {Somsen}, \citenamefont {Bei}, \citenamefont {Eggeler},\ and\ \citenamefont {George}}]{otto2013influences}%
  \BibitemOpen
  \bibfield  {author} {\bibinfo {author} {\bibfnamefont {F.}~\bibnamefont {Otto}}, \bibinfo {author} {\bibfnamefont {A.}~\bibnamefont {Dlouh{\`y}}}, \bibinfo {author} {\bibfnamefont {C.}~\bibnamefont {Somsen}}, \bibinfo {author} {\bibfnamefont {H.}~\bibnamefont {Bei}}, \bibinfo {author} {\bibfnamefont {G.}~\bibnamefont {Eggeler}},\ and\ \bibinfo {author} {\bibfnamefont {E.~P.}\ \bibnamefont {George}},\ }\bibfield  {title} {\bibinfo {title} {The influences of temperature and microstructure on the tensile properties of a cocrfemnni high-entropy alloy},\ }\href@noop {} {\bibfield  {journal} {\bibinfo  {journal} {Acta Materialia}\ }\textbf {\bibinfo {volume} {61}},\ \bibinfo {pages} {5743} (\bibinfo {year} {2013})}\BibitemShut {NoStop}%
\bibitem [{\citenamefont {Mo}\ \emph {et~al.}(2012)\citenamefont {Mo}, \citenamefont {Ong},\ and\ \citenamefont {Ceder}}]{mo2012first}%
  \BibitemOpen
  \bibfield  {author} {\bibinfo {author} {\bibfnamefont {Y.}~\bibnamefont {Mo}}, \bibinfo {author} {\bibfnamefont {S.~P.}\ \bibnamefont {Ong}},\ and\ \bibinfo {author} {\bibfnamefont {G.}~\bibnamefont {Ceder}},\ }\bibfield  {title} {\bibinfo {title} {First principles study of the li10gep2s12 lithium super ionic conductor material},\ }\href@noop {} {\bibfield  {journal} {\bibinfo  {journal} {Chemistry of Materials}\ }\textbf {\bibinfo {volume} {24}},\ \bibinfo {pages} {15} (\bibinfo {year} {2012})}\BibitemShut {NoStop}%
\bibitem [{\citenamefont {Mei}\ \emph {et~al.}(1993)\citenamefont {Mei}, \citenamefont {Wang}, \citenamefont {Chen},\ and\ \citenamefont {Wu}}]{mei1993nonlinear}%
  \BibitemOpen
  \bibfield  {author} {\bibinfo {author} {\bibfnamefont {L.}~\bibnamefont {Mei}}, \bibinfo {author} {\bibfnamefont {Y.}~\bibnamefont {Wang}}, \bibinfo {author} {\bibfnamefont {C.}~\bibnamefont {Chen}},\ and\ \bibinfo {author} {\bibfnamefont {B.}~\bibnamefont {Wu}},\ }\bibfield  {title} {\bibinfo {title} {Nonlinear optical materials based on mbe2bo3f2 (m= na, k)},\ }\href@noop {} {\bibfield  {journal} {\bibinfo  {journal} {Journal of applied physics}\ }\textbf {\bibinfo {volume} {74}},\ \bibinfo {pages} {7014} (\bibinfo {year} {1993})}\BibitemShut {NoStop}%
\bibitem [{\citenamefont {Wu}\ \emph {et~al.}(1996)\citenamefont {Wu}, \citenamefont {Tang}, \citenamefont {Ye},\ and\ \citenamefont {Chen}}]{wu1996linear}%
  \BibitemOpen
  \bibfield  {author} {\bibinfo {author} {\bibfnamefont {B.}~\bibnamefont {Wu}}, \bibinfo {author} {\bibfnamefont {D.}~\bibnamefont {Tang}}, \bibinfo {author} {\bibfnamefont {N.}~\bibnamefont {Ye}},\ and\ \bibinfo {author} {\bibfnamefont {C.}~\bibnamefont {Chen}},\ }\bibfield  {title} {\bibinfo {title} {Linear and nonlinear optical properties of the kbe2bo3f2 (kbbf) crystal},\ }\href@noop {} {\bibfield  {journal} {\bibinfo  {journal} {Optical Materials}\ }\textbf {\bibinfo {volume} {5}},\ \bibinfo {pages} {105} (\bibinfo {year} {1996})}\BibitemShut {NoStop}%
\bibitem [{\citenamefont {Xie}\ \emph {et~al.}(2023)\citenamefont {Xie}, \citenamefont {Tikhonov}, \citenamefont {Chu}, \citenamefont {Wu}, \citenamefont {Kruglov}, \citenamefont {Pan},\ and\ \citenamefont {Yang}}]{xie2023prediction}%
  \BibitemOpen
  \bibfield  {author} {\bibinfo {author} {\bibfnamefont {C.}~\bibnamefont {Xie}}, \bibinfo {author} {\bibfnamefont {E.}~\bibnamefont {Tikhonov}}, \bibinfo {author} {\bibfnamefont {D.}~\bibnamefont {Chu}}, \bibinfo {author} {\bibfnamefont {M.}~\bibnamefont {Wu}}, \bibinfo {author} {\bibfnamefont {I.}~\bibnamefont {Kruglov}}, \bibinfo {author} {\bibfnamefont {S.}~\bibnamefont {Pan}},\ and\ \bibinfo {author} {\bibfnamefont {Z.}~\bibnamefont {Yang}},\ }\bibfield  {title} {\bibinfo {title} {A prediction-driven database to enable rapid discovery of nonlinear optical materials},\ }\href@noop {} {\bibfield  {journal} {\bibinfo  {journal} {Science China Materials}\ }\textbf {\bibinfo {volume} {66}},\ \bibinfo {pages} {4473} (\bibinfo {year} {2023})}\BibitemShut {NoStop}%
\bibitem [{\citenamefont {Park}\ \emph {et~al.}(2025)\citenamefont {Park}, \citenamefont {Onwuli}, \citenamefont {Butler},\ and\ \citenamefont {Walsh}}]{park2025mapping}%
  \BibitemOpen
  \bibfield  {author} {\bibinfo {author} {\bibfnamefont {H.}~\bibnamefont {Park}}, \bibinfo {author} {\bibfnamefont {A.}~\bibnamefont {Onwuli}}, \bibinfo {author} {\bibfnamefont {K.~T.}\ \bibnamefont {Butler}},\ and\ \bibinfo {author} {\bibfnamefont {A.}~\bibnamefont {Walsh}},\ }\bibfield  {title} {\bibinfo {title} {Mapping inorganic crystal chemical space},\ }\href@noop {} {\bibfield  {journal} {\bibinfo  {journal} {Faraday Discussions}\ }\textbf {\bibinfo {volume} {256}},\ \bibinfo {pages} {601} (\bibinfo {year} {2025})}\BibitemShut {NoStop}%
\bibitem [{\citenamefont {Pickard}\ and\ \citenamefont {Needs}(2011)}]{Pickard_2011}%
  \BibitemOpen
  \bibfield  {author} {\bibinfo {author} {\bibfnamefont {C.~J.}\ \bibnamefont {Pickard}}\ and\ \bibinfo {author} {\bibfnamefont {R.~J.}\ \bibnamefont {Needs}},\ }\bibfield  {title} {\bibinfo {title} {Ab initio random structure searching},\ }\href {https://doi.org/10.1088/0953-8984/23/5/053201} {\bibfield  {journal} {\bibinfo  {journal} {Journal of Physics: Condensed Matter}\ }\textbf {\bibinfo {volume} {23}},\ \bibinfo {pages} {053201} (\bibinfo {year} {2011})}\BibitemShut {NoStop}%
\bibitem [{\citenamefont {Wang}\ \emph {et~al.}(2010)\citenamefont {Wang}, \citenamefont {Lv}, \citenamefont {Zhu},\ and\ \citenamefont {Ma}}]{wang2010crystal}%
  \BibitemOpen
  \bibfield  {author} {\bibinfo {author} {\bibfnamefont {Y.}~\bibnamefont {Wang}}, \bibinfo {author} {\bibfnamefont {J.}~\bibnamefont {Lv}}, \bibinfo {author} {\bibfnamefont {L.}~\bibnamefont {Zhu}},\ and\ \bibinfo {author} {\bibfnamefont {Y.}~\bibnamefont {Ma}},\ }\bibfield  {title} {\bibinfo {title} {Crystal structure prediction via particle-swarm optimization},\ }\href@noop {} {\bibfield  {journal} {\bibinfo  {journal} {Physical Review B—Condensed Matter and Materials Physics}\ }\textbf {\bibinfo {volume} {82}},\ \bibinfo {pages} {094116} (\bibinfo {year} {2010})}\BibitemShut {NoStop}%
\bibitem [{\citenamefont {Wang}\ \emph {et~al.}(2012{\natexlab{a}})\citenamefont {Wang}, \citenamefont {Lv}, \citenamefont {Zhu},\ and\ \citenamefont {Ma}}]{wang2012calypso}%
  \BibitemOpen
  \bibfield  {author} {\bibinfo {author} {\bibfnamefont {Y.}~\bibnamefont {Wang}}, \bibinfo {author} {\bibfnamefont {J.}~\bibnamefont {Lv}}, \bibinfo {author} {\bibfnamefont {L.}~\bibnamefont {Zhu}},\ and\ \bibinfo {author} {\bibfnamefont {Y.}~\bibnamefont {Ma}},\ }\bibfield  {title} {\bibinfo {title} {Calypso: A method for crystal structure prediction},\ }\href@noop {} {\bibfield  {journal} {\bibinfo  {journal} {Computer Physics Communications}\ }\textbf {\bibinfo {volume} {183}},\ \bibinfo {pages} {2063} (\bibinfo {year} {2012}{\natexlab{a}})}\BibitemShut {NoStop}%
\bibitem [{\citenamefont {Oganov}\ and\ \citenamefont {Glass}(2006)}]{oganov2006crystal}%
  \BibitemOpen
  \bibfield  {author} {\bibinfo {author} {\bibfnamefont {A.~R.}\ \bibnamefont {Oganov}}\ and\ \bibinfo {author} {\bibfnamefont {C.~W.}\ \bibnamefont {Glass}},\ }\bibfield  {title} {\bibinfo {title} {Crystal structure prediction using ab initio evolutionary techniques: Principles and applications},\ }\href@noop {} {\bibfield  {journal} {\bibinfo  {journal} {The Journal of chemical physics}\ }\textbf {\bibinfo {volume} {124}} (\bibinfo {year} {2006})}\BibitemShut {NoStop}%
\bibitem [{\citenamefont {Behler}\ and\ \citenamefont {Parrinello}(2007)}]{behler2007generalized}%
  \BibitemOpen
  \bibfield  {author} {\bibinfo {author} {\bibfnamefont {J.}~\bibnamefont {Behler}}\ and\ \bibinfo {author} {\bibfnamefont {M.}~\bibnamefont {Parrinello}},\ }\bibfield  {title} {\bibinfo {title} {Generalized neural-network representation of high-dimensional potential-energy surfaces},\ }\href@noop {} {\bibfield  {journal} {\bibinfo  {journal} {Physical review letters}\ }\textbf {\bibinfo {volume} {98}},\ \bibinfo {pages} {146401} (\bibinfo {year} {2007})}\BibitemShut {NoStop}%
\bibitem [{\citenamefont {Shapeev}(2016)}]{shapeev2016moment}%
  \BibitemOpen
  \bibfield  {author} {\bibinfo {author} {\bibfnamefont {A.~V.}\ \bibnamefont {Shapeev}},\ }\bibfield  {title} {\bibinfo {title} {Moment tensor potentials: A class of systematically improvable interatomic potentials},\ }\href@noop {} {\bibfield  {journal} {\bibinfo  {journal} {Multiscale Modeling \& Simulation}\ }\textbf {\bibinfo {volume} {14}},\ \bibinfo {pages} {1153} (\bibinfo {year} {2016})}\BibitemShut {NoStop}%
\bibitem [{\citenamefont {Bart{\'o}k}\ \emph {et~al.}(2010)\citenamefont {Bart{\'o}k}, \citenamefont {Payne}, \citenamefont {Kondor},\ and\ \citenamefont {Cs{\'a}nyi}}]{bartok2010gaussian}%
  \BibitemOpen
  \bibfield  {author} {\bibinfo {author} {\bibfnamefont {A.~P.}\ \bibnamefont {Bart{\'o}k}}, \bibinfo {author} {\bibfnamefont {M.~C.}\ \bibnamefont {Payne}}, \bibinfo {author} {\bibfnamefont {R.}~\bibnamefont {Kondor}},\ and\ \bibinfo {author} {\bibfnamefont {G.}~\bibnamefont {Cs{\'a}nyi}},\ }\bibfield  {title} {\bibinfo {title} {Gaussian approximation potentials: The accuracy of quantum mechanics, without the electrons},\ }\href@noop {} {\bibfield  {journal} {\bibinfo  {journal} {Physical review letters}\ }\textbf {\bibinfo {volume} {104}},\ \bibinfo {pages} {136403} (\bibinfo {year} {2010})}\BibitemShut {NoStop}%
\bibitem [{\citenamefont {Zhang}\ \emph {et~al.}(2018)\citenamefont {Zhang}, \citenamefont {Han}, \citenamefont {Wang}, \citenamefont {Saidi}, \citenamefont {Car} \emph {et~al.}}]{zhang2018end}%
  \BibitemOpen
  \bibfield  {author} {\bibinfo {author} {\bibfnamefont {L.}~\bibnamefont {Zhang}}, \bibinfo {author} {\bibfnamefont {J.}~\bibnamefont {Han}}, \bibinfo {author} {\bibfnamefont {H.}~\bibnamefont {Wang}}, \bibinfo {author} {\bibfnamefont {W.}~\bibnamefont {Saidi}}, \bibinfo {author} {\bibfnamefont {R.}~\bibnamefont {Car}}, \emph {et~al.},\ }\bibfield  {title} {\bibinfo {title} {End-to-end symmetry preserving inter-atomic potential energy model for finite and extended systems},\ }\href@noop {} {\bibfield  {journal} {\bibinfo  {journal} {Advances in neural information processing systems}\ }\textbf {\bibinfo {volume} {31}} (\bibinfo {year} {2018})}\BibitemShut {NoStop}%
\bibitem [{\citenamefont {Chen}\ and\ \citenamefont {Ong}(2022)}]{chen2022universal}%
  \BibitemOpen
  \bibfield  {author} {\bibinfo {author} {\bibfnamefont {C.}~\bibnamefont {Chen}}\ and\ \bibinfo {author} {\bibfnamefont {S.~P.}\ \bibnamefont {Ong}},\ }\bibfield  {title} {\bibinfo {title} {A universal graph deep learning interatomic potential for the periodic table},\ }\href@noop {} {\bibfield  {journal} {\bibinfo  {journal} {Nature Computational Science}\ }\textbf {\bibinfo {volume} {2}},\ \bibinfo {pages} {718} (\bibinfo {year} {2022})}\BibitemShut {NoStop}%
\bibitem [{\citenamefont {Batatia}\ \emph {et~al.}(2022)\citenamefont {Batatia}, \citenamefont {Kovacs}, \citenamefont {Simm}, \citenamefont {Ortner},\ and\ \citenamefont {Cs{\'a}nyi}}]{batatia2022mace}%
  \BibitemOpen
  \bibfield  {author} {\bibinfo {author} {\bibfnamefont {I.}~\bibnamefont {Batatia}}, \bibinfo {author} {\bibfnamefont {D.~P.}\ \bibnamefont {Kovacs}}, \bibinfo {author} {\bibfnamefont {G.}~\bibnamefont {Simm}}, \bibinfo {author} {\bibfnamefont {C.}~\bibnamefont {Ortner}},\ and\ \bibinfo {author} {\bibfnamefont {G.}~\bibnamefont {Cs{\'a}nyi}},\ }\bibfield  {title} {\bibinfo {title} {Mace: Higher order equivariant message passing neural networks for fast and accurate force fields},\ }\href@noop {} {\bibfield  {journal} {\bibinfo  {journal} {Advances in Neural Information Processing Systems}\ }\textbf {\bibinfo {volume} {35}},\ \bibinfo {pages} {11423} (\bibinfo {year} {2022})}\BibitemShut {NoStop}%
\bibitem [{\citenamefont {Juelsholt}(2025)}]{juelsholt2025continued}%
  \BibitemOpen
  \bibfield  {author} {\bibinfo {author} {\bibfnamefont {M.}~\bibnamefont {Juelsholt}},\ }\bibfield  {title} {\bibinfo {title} {Continued challenges in high-throughput materials predictions: Mattergen predicts compounds from the training dataset.},\ }\href@noop {} {\bibfield  {journal} {\bibinfo  {journal} {chemrxiv-2025-mkls8}\ } (\bibinfo {year} {2025})}\BibitemShut {NoStop}%
\bibitem [{\citenamefont {Wang}\ \emph {et~al.}(2024)\citenamefont {Wang}, \citenamefont {Wang}, \citenamefont {Luo}, \citenamefont {Gao}, \citenamefont {Sun}, \citenamefont {Lv}, \citenamefont {Wang}, \citenamefont {Wang},\ and\ \citenamefont {Ma}}]{wang2024concurrent}%
  \BibitemOpen
  \bibfield  {author} {\bibinfo {author} {\bibfnamefont {Z.}~\bibnamefont {Wang}}, \bibinfo {author} {\bibfnamefont {X.}~\bibnamefont {Wang}}, \bibinfo {author} {\bibfnamefont {X.}~\bibnamefont {Luo}}, \bibinfo {author} {\bibfnamefont {P.}~\bibnamefont {Gao}}, \bibinfo {author} {\bibfnamefont {Y.}~\bibnamefont {Sun}}, \bibinfo {author} {\bibfnamefont {J.}~\bibnamefont {Lv}}, \bibinfo {author} {\bibfnamefont {H.}~\bibnamefont {Wang}}, \bibinfo {author} {\bibfnamefont {Y.}~\bibnamefont {Wang}},\ and\ \bibinfo {author} {\bibfnamefont {Y.}~\bibnamefont {Ma}},\ }\bibfield  {title} {\bibinfo {title} {Concurrent learning scheme for crystal structure prediction},\ }\href@noop {} {\bibfield  {journal} {\bibinfo  {journal} {Physical Review B}\ }\textbf {\bibinfo {volume} {109}},\ \bibinfo {pages} {094117} (\bibinfo {year} {2024})}\BibitemShut {NoStop}%
\bibitem [{\citenamefont {Tong}\ \emph {et~al.}(2018)\citenamefont {Tong}, \citenamefont {Xue}, \citenamefont {Lv}, \citenamefont {Wang},\ and\ \citenamefont {Ma}}]{tong2018accelerating}%
  \BibitemOpen
  \bibfield  {author} {\bibinfo {author} {\bibfnamefont {Q.}~\bibnamefont {Tong}}, \bibinfo {author} {\bibfnamefont {L.}~\bibnamefont {Xue}}, \bibinfo {author} {\bibfnamefont {J.}~\bibnamefont {Lv}}, \bibinfo {author} {\bibfnamefont {Y.}~\bibnamefont {Wang}},\ and\ \bibinfo {author} {\bibfnamefont {Y.}~\bibnamefont {Ma}},\ }\bibfield  {title} {\bibinfo {title} {Accelerating calypso structure prediction by data-driven learning of a potential energy surface},\ }\href@noop {} {\bibfield  {journal} {\bibinfo  {journal} {Faraday discussions}\ }\textbf {\bibinfo {volume} {211}},\ \bibinfo {pages} {31} (\bibinfo {year} {2018})}\BibitemShut {NoStop}%
\bibitem [{\citenamefont {Tong}\ \emph {et~al.}(2020)\citenamefont {Tong}, \citenamefont {Gao}, \citenamefont {Liu}, \citenamefont {Xie}, \citenamefont {Lv}, \citenamefont {Wang},\ and\ \citenamefont {Zhao}}]{tong2020combining}%
  \BibitemOpen
  \bibfield  {author} {\bibinfo {author} {\bibfnamefont {Q.}~\bibnamefont {Tong}}, \bibinfo {author} {\bibfnamefont {P.}~\bibnamefont {Gao}}, \bibinfo {author} {\bibfnamefont {H.}~\bibnamefont {Liu}}, \bibinfo {author} {\bibfnamefont {Y.}~\bibnamefont {Xie}}, \bibinfo {author} {\bibfnamefont {J.}~\bibnamefont {Lv}}, \bibinfo {author} {\bibfnamefont {Y.}~\bibnamefont {Wang}},\ and\ \bibinfo {author} {\bibfnamefont {J.}~\bibnamefont {Zhao}},\ }\bibfield  {title} {\bibinfo {title} {Combining machine learning potential and structure prediction for accelerated materials design and discovery},\ }\href@noop {} {\bibfield  {journal} {\bibinfo  {journal} {The journal of physical chemistry letters}\ }\textbf {\bibinfo {volume} {11}},\ \bibinfo {pages} {8710} (\bibinfo {year} {2020})}\BibitemShut {NoStop}%
\bibitem [{\citenamefont {Curtarolo}\ \emph {et~al.}(2012)\citenamefont {Curtarolo}, \citenamefont {Setyawan}, \citenamefont {Hart}, \citenamefont {Jahnatek}, \citenamefont {Chepulskii}, \citenamefont {Taylor}, \citenamefont {Wang}, \citenamefont {Xue}, \citenamefont {Yang}, \citenamefont {Levy} \emph {et~al.}}]{curtarolo2012aflow}%
  \BibitemOpen
  \bibfield  {author} {\bibinfo {author} {\bibfnamefont {S.}~\bibnamefont {Curtarolo}}, \bibinfo {author} {\bibfnamefont {W.}~\bibnamefont {Setyawan}}, \bibinfo {author} {\bibfnamefont {G.~L.}\ \bibnamefont {Hart}}, \bibinfo {author} {\bibfnamefont {M.}~\bibnamefont {Jahnatek}}, \bibinfo {author} {\bibfnamefont {R.~V.}\ \bibnamefont {Chepulskii}}, \bibinfo {author} {\bibfnamefont {R.~H.}\ \bibnamefont {Taylor}}, \bibinfo {author} {\bibfnamefont {S.}~\bibnamefont {Wang}}, \bibinfo {author} {\bibfnamefont {J.}~\bibnamefont {Xue}}, \bibinfo {author} {\bibfnamefont {K.}~\bibnamefont {Yang}}, \bibinfo {author} {\bibfnamefont {O.}~\bibnamefont {Levy}}, \emph {et~al.},\ }\bibfield  {title} {\bibinfo {title} {Aflow: An automatic framework for high-throughput materials discovery},\ }\href@noop {} {\bibfield  {journal} {\bibinfo  {journal} {Computational Materials Science}\ }\textbf {\bibinfo {volume} {58}},\ \bibinfo {pages} {218} (\bibinfo {year} {2012})}\BibitemShut {NoStop}%
\bibitem [{\citenamefont {Pizzi}\ \emph {et~al.}(2016)\citenamefont {Pizzi}, \citenamefont {Cepellotti}, \citenamefont {Sabatini}, \citenamefont {Marzari},\ and\ \citenamefont {Kozinsky}}]{pizzi2016aiida}%
  \BibitemOpen
  \bibfield  {author} {\bibinfo {author} {\bibfnamefont {G.}~\bibnamefont {Pizzi}}, \bibinfo {author} {\bibfnamefont {A.}~\bibnamefont {Cepellotti}}, \bibinfo {author} {\bibfnamefont {R.}~\bibnamefont {Sabatini}}, \bibinfo {author} {\bibfnamefont {N.}~\bibnamefont {Marzari}},\ and\ \bibinfo {author} {\bibfnamefont {B.}~\bibnamefont {Kozinsky}},\ }\bibfield  {title} {\bibinfo {title} {Aiida: automated interactive infrastructure and database for computational science},\ }\href@noop {} {\bibfield  {journal} {\bibinfo  {journal} {Computational Materials Science}\ }\textbf {\bibinfo {volume} {111}},\ \bibinfo {pages} {218} (\bibinfo {year} {2016})}\BibitemShut {NoStop}%
\bibitem [{\citenamefont {Batzner}\ \emph {et~al.}(2022)\citenamefont {Batzner}, \citenamefont {Musaelian}, \citenamefont {Sun}, \citenamefont {Geiger}, \citenamefont {Mailoa}, \citenamefont {Kornbluth}, \citenamefont {Molinari}, \citenamefont {Smidt},\ and\ \citenamefont {Kozinsky}}]{batzner20223}%
  \BibitemOpen
  \bibfield  {author} {\bibinfo {author} {\bibfnamefont {S.}~\bibnamefont {Batzner}}, \bibinfo {author} {\bibfnamefont {A.}~\bibnamefont {Musaelian}}, \bibinfo {author} {\bibfnamefont {L.}~\bibnamefont {Sun}}, \bibinfo {author} {\bibfnamefont {M.}~\bibnamefont {Geiger}}, \bibinfo {author} {\bibfnamefont {J.~P.}\ \bibnamefont {Mailoa}}, \bibinfo {author} {\bibfnamefont {M.}~\bibnamefont {Kornbluth}}, \bibinfo {author} {\bibfnamefont {N.}~\bibnamefont {Molinari}}, \bibinfo {author} {\bibfnamefont {T.~E.}\ \bibnamefont {Smidt}},\ and\ \bibinfo {author} {\bibfnamefont {B.}~\bibnamefont {Kozinsky}},\ }\bibfield  {title} {\bibinfo {title} {E (3)-equivariant graph neural networks for data-efficient and accurate interatomic potentials},\ }\href@noop {} {\bibfield  {journal} {\bibinfo  {journal} {Nature communications}\ }\textbf {\bibinfo {volume} {13}},\ \bibinfo {pages} {2453} (\bibinfo {year} {2022})}\BibitemShut {NoStop}%
\bibitem [{\citenamefont {Yu}\ \emph {et~al.}(2024)\citenamefont {Yu}, \citenamefont {Giantomassi}, \citenamefont {Materzanini}, \citenamefont {Wang},\ and\ \citenamefont {Rignanese}}]{yu2024systematic}%
  \BibitemOpen
  \bibfield  {author} {\bibinfo {author} {\bibfnamefont {H.}~\bibnamefont {Yu}}, \bibinfo {author} {\bibfnamefont {M.}~\bibnamefont {Giantomassi}}, \bibinfo {author} {\bibfnamefont {G.}~\bibnamefont {Materzanini}}, \bibinfo {author} {\bibfnamefont {J.}~\bibnamefont {Wang}},\ and\ \bibinfo {author} {\bibfnamefont {G.-M.}\ \bibnamefont {Rignanese}},\ }\bibfield  {title} {\bibinfo {title} {Systematic assessment of various universal machine-learning interatomic potentials},\ }\href@noop {} {\bibfield  {journal} {\bibinfo  {journal} {Materials Genome Engineering Advances}\ }\textbf {\bibinfo {volume} {2}},\ \bibinfo {pages} {e58} (\bibinfo {year} {2024})}\BibitemShut {NoStop}%
\bibitem [{\citenamefont {Drautz}(2019)}]{drautz2019atomic}%
  \BibitemOpen
  \bibfield  {author} {\bibinfo {author} {\bibfnamefont {R.}~\bibnamefont {Drautz}},\ }\bibfield  {title} {\bibinfo {title} {Atomic cluster expansion for accurate and transferable interatomic potentials},\ }\href@noop {} {\bibfield  {journal} {\bibinfo  {journal} {Physical Review B}\ }\textbf {\bibinfo {volume} {99}},\ \bibinfo {pages} {014104} (\bibinfo {year} {2019})}\BibitemShut {NoStop}%
\bibitem [{\citenamefont {Fan}\ \emph {et~al.}(2022)\citenamefont {Fan}, \citenamefont {Wang}, \citenamefont {Ying}, \citenamefont {Song}, \citenamefont {Wang}, \citenamefont {Wang}, \citenamefont {Zeng}, \citenamefont {Xu}, \citenamefont {Lindgren}, \citenamefont {Rahm} \emph {et~al.}}]{fan2022gpumd}%
  \BibitemOpen
  \bibfield  {author} {\bibinfo {author} {\bibfnamefont {Z.}~\bibnamefont {Fan}}, \bibinfo {author} {\bibfnamefont {Y.}~\bibnamefont {Wang}}, \bibinfo {author} {\bibfnamefont {P.}~\bibnamefont {Ying}}, \bibinfo {author} {\bibfnamefont {K.}~\bibnamefont {Song}}, \bibinfo {author} {\bibfnamefont {J.}~\bibnamefont {Wang}}, \bibinfo {author} {\bibfnamefont {Y.}~\bibnamefont {Wang}}, \bibinfo {author} {\bibfnamefont {Z.}~\bibnamefont {Zeng}}, \bibinfo {author} {\bibfnamefont {K.}~\bibnamefont {Xu}}, \bibinfo {author} {\bibfnamefont {E.}~\bibnamefont {Lindgren}}, \bibinfo {author} {\bibfnamefont {J.~M.}\ \bibnamefont {Rahm}}, \emph {et~al.},\ }\bibfield  {title} {\bibinfo {title} {Gpumd: A package for constructing accurate machine-learned potentials and performing highly efficient atomistic simulations},\ }\href@noop {} {\bibfield  {journal} {\bibinfo  {journal} {The Journal of Chemical Physics}\ }\textbf {\bibinfo {volume} {157}} (\bibinfo {year} {2022})}\BibitemShut {NoStop}%
\bibitem [{\citenamefont {Vaswani}(2017)}]{vaswani2017attention}%
  \BibitemOpen
  \bibfield  {author} {\bibinfo {author} {\bibfnamefont {A.}~\bibnamefont {Vaswani}},\ }\bibfield  {title} {\bibinfo {title} {Attention is all you need},\ }\href@noop {} {\bibfield  {journal} {\bibinfo  {journal} {Advances in Neural Information Processing Systems}\ } (\bibinfo {year} {2017})}\BibitemShut {NoStop}%
\bibitem [{\citenamefont {Gilmer}\ \emph {et~al.}(2017)\citenamefont {Gilmer}, \citenamefont {Schoenholz}, \citenamefont {Riley}, \citenamefont {Vinyals},\ and\ \citenamefont {Dahl}}]{gilmer2017neural}%
  \BibitemOpen
  \bibfield  {author} {\bibinfo {author} {\bibfnamefont {J.}~\bibnamefont {Gilmer}}, \bibinfo {author} {\bibfnamefont {S.~S.}\ \bibnamefont {Schoenholz}}, \bibinfo {author} {\bibfnamefont {P.~F.}\ \bibnamefont {Riley}}, \bibinfo {author} {\bibfnamefont {O.}~\bibnamefont {Vinyals}},\ and\ \bibinfo {author} {\bibfnamefont {G.~E.}\ \bibnamefont {Dahl}},\ }\bibfield  {title} {\bibinfo {title} {Neural message passing for quantum chemistry},\ }in\ \href@noop {} {\emph {\bibinfo {booktitle} {International conference on machine learning}}}\ (\bibinfo {organization} {PMLR},\ \bibinfo {year} {2017})\ pp.\ \bibinfo {pages} {1263--1272}\BibitemShut {NoStop}%
\bibitem [{\citenamefont {R{\"o}cken}\ and\ \citenamefont {Zavadlav}(2024)}]{rocken2024accurate}%
  \BibitemOpen
  \bibfield  {author} {\bibinfo {author} {\bibfnamefont {S.}~\bibnamefont {R{\"o}cken}}\ and\ \bibinfo {author} {\bibfnamefont {J.}~\bibnamefont {Zavadlav}},\ }\bibfield  {title} {\bibinfo {title} {Accurate machine learning force fields via experimental and simulation data fusion},\ }\href@noop {} {\bibfield  {journal} {\bibinfo  {journal} {npj Computational Materials}\ }\textbf {\bibinfo {volume} {10}},\ \bibinfo {pages} {69} (\bibinfo {year} {2024})}\BibitemShut {NoStop}%
\bibitem [{\citenamefont {Liu}\ \emph {et~al.}(2021)\citenamefont {Liu}, \citenamefont {Niu},\ and\ \citenamefont {Oganov}}]{liu2021copex}%
  \BibitemOpen
  \bibfield  {author} {\bibinfo {author} {\bibfnamefont {X.}~\bibnamefont {Liu}}, \bibinfo {author} {\bibfnamefont {H.}~\bibnamefont {Niu}},\ and\ \bibinfo {author} {\bibfnamefont {A.~R.}\ \bibnamefont {Oganov}},\ }\bibfield  {title} {\bibinfo {title} {Copex: co-evolutionary crystal structure prediction algorithm for complex systems},\ }\href@noop {} {\bibfield  {journal} {\bibinfo  {journal} {npj Computational Materials}\ }\textbf {\bibinfo {volume} {7}},\ \bibinfo {pages} {199} (\bibinfo {year} {2021})}\BibitemShut {NoStop}%
\bibitem [{\citenamefont {Wang}\ \emph {et~al.}(2012{\natexlab{b}})\citenamefont {Wang}, \citenamefont {Tse}, \citenamefont {Tanaka}, \citenamefont {Iitaka},\ and\ \citenamefont {Ma}}]{wang2012superconductive}%
  \BibitemOpen
  \bibfield  {author} {\bibinfo {author} {\bibfnamefont {H.}~\bibnamefont {Wang}}, \bibinfo {author} {\bibfnamefont {J.~S.}\ \bibnamefont {Tse}}, \bibinfo {author} {\bibfnamefont {K.}~\bibnamefont {Tanaka}}, \bibinfo {author} {\bibfnamefont {T.}~\bibnamefont {Iitaka}},\ and\ \bibinfo {author} {\bibfnamefont {Y.}~\bibnamefont {Ma}},\ }\bibfield  {title} {\bibinfo {title} {Superconductive sodalite-like clathrate calcium hydride at high pressures},\ }\href@noop {} {\bibfield  {journal} {\bibinfo  {journal} {Proceedings of the National Academy of Sciences}\ }\textbf {\bibinfo {volume} {109}},\ \bibinfo {pages} {6463} (\bibinfo {year} {2012}{\natexlab{b}})}\BibitemShut {NoStop}%
\bibitem [{\citenamefont {Ma}\ \emph {et~al.}(2022)\citenamefont {Ma}, \citenamefont {Wang}, \citenamefont {Xie}, \citenamefont {Yang}, \citenamefont {Wang}, \citenamefont {Zhou}, \citenamefont {Liu}, \citenamefont {Yu}, \citenamefont {Zhao}, \citenamefont {Wang} \emph {et~al.}}]{ma2022high}%
  \BibitemOpen
  \bibfield  {author} {\bibinfo {author} {\bibfnamefont {L.}~\bibnamefont {Ma}}, \bibinfo {author} {\bibfnamefont {K.}~\bibnamefont {Wang}}, \bibinfo {author} {\bibfnamefont {Y.}~\bibnamefont {Xie}}, \bibinfo {author} {\bibfnamefont {X.}~\bibnamefont {Yang}}, \bibinfo {author} {\bibfnamefont {Y.}~\bibnamefont {Wang}}, \bibinfo {author} {\bibfnamefont {M.}~\bibnamefont {Zhou}}, \bibinfo {author} {\bibfnamefont {H.}~\bibnamefont {Liu}}, \bibinfo {author} {\bibfnamefont {X.}~\bibnamefont {Yu}}, \bibinfo {author} {\bibfnamefont {Y.}~\bibnamefont {Zhao}}, \bibinfo {author} {\bibfnamefont {H.}~\bibnamefont {Wang}}, \emph {et~al.},\ }\bibfield  {title} {\bibinfo {title} {High-temperature superconducting phase in clathrate calcium hydride cah 6 up to 215 k at a pressure of 172 gpa},\ }\href@noop {} {\bibfield  {journal} {\bibinfo  {journal} {Physical Review Letters}\ }\textbf {\bibinfo {volume} {128}},\ \bibinfo {pages} {167001} (\bibinfo {year} {2022})}\BibitemShut {NoStop}%
\bibitem [{\citenamefont {Peng}\ \emph {et~al.}(2017)\citenamefont {Peng}, \citenamefont {Sun}, \citenamefont {Pickard}, \citenamefont {Needs}, \citenamefont {Wu},\ and\ \citenamefont {Ma}}]{peng2017hydrogen}%
  \BibitemOpen
  \bibfield  {author} {\bibinfo {author} {\bibfnamefont {F.}~\bibnamefont {Peng}}, \bibinfo {author} {\bibfnamefont {Y.}~\bibnamefont {Sun}}, \bibinfo {author} {\bibfnamefont {C.~J.}\ \bibnamefont {Pickard}}, \bibinfo {author} {\bibfnamefont {R.~J.}\ \bibnamefont {Needs}}, \bibinfo {author} {\bibfnamefont {Q.}~\bibnamefont {Wu}},\ and\ \bibinfo {author} {\bibfnamefont {Y.}~\bibnamefont {Ma}},\ }\bibfield  {title} {\bibinfo {title} {Hydrogen clathrate structures in rare earth hydrides at high pressures: possible route to room-temperature superconductivity},\ }\href@noop {} {\bibfield  {journal} {\bibinfo  {journal} {Physical review letters}\ }\textbf {\bibinfo {volume} {119}},\ \bibinfo {pages} {107001} (\bibinfo {year} {2017})}\BibitemShut {NoStop}%
\bibitem [{\citenamefont {Liu}\ \emph {et~al.}(2017)\citenamefont {Liu}, \citenamefont {Naumov}, \citenamefont {Hoffmann}, \citenamefont {Ashcroft},\ and\ \citenamefont {Hemley}}]{liu2017potential}%
  \BibitemOpen
  \bibfield  {author} {\bibinfo {author} {\bibfnamefont {H.}~\bibnamefont {Liu}}, \bibinfo {author} {\bibfnamefont {I.~I.}\ \bibnamefont {Naumov}}, \bibinfo {author} {\bibfnamefont {R.}~\bibnamefont {Hoffmann}}, \bibinfo {author} {\bibfnamefont {N.}~\bibnamefont {Ashcroft}},\ and\ \bibinfo {author} {\bibfnamefont {R.~J.}\ \bibnamefont {Hemley}},\ }\bibfield  {title} {\bibinfo {title} {Potential high-tc superconducting lanthanum and yttrium hydrides at high pressure},\ }\href@noop {} {\bibfield  {journal} {\bibinfo  {journal} {Proceedings of the National Academy of Sciences}\ }\textbf {\bibinfo {volume} {114}},\ \bibinfo {pages} {6990} (\bibinfo {year} {2017})}\BibitemShut {NoStop}%
\bibitem [{\citenamefont {Drozdov}\ \emph {et~al.}(2019)\citenamefont {Drozdov}, \citenamefont {Kong}, \citenamefont {Minkov}, \citenamefont {Besedin}, \citenamefont {Kuzovnikov}, \citenamefont {Mozaffari}, \citenamefont {Balicas}, \citenamefont {Balakirev}, \citenamefont {Graf}, \citenamefont {Prakapenka} \emph {et~al.}}]{drozdov2019superconductivity}%
  \BibitemOpen
  \bibfield  {author} {\bibinfo {author} {\bibfnamefont {A.}~\bibnamefont {Drozdov}}, \bibinfo {author} {\bibfnamefont {P.}~\bibnamefont {Kong}}, \bibinfo {author} {\bibfnamefont {V.}~\bibnamefont {Minkov}}, \bibinfo {author} {\bibfnamefont {S.}~\bibnamefont {Besedin}}, \bibinfo {author} {\bibfnamefont {M.}~\bibnamefont {Kuzovnikov}}, \bibinfo {author} {\bibfnamefont {S.}~\bibnamefont {Mozaffari}}, \bibinfo {author} {\bibfnamefont {L.}~\bibnamefont {Balicas}}, \bibinfo {author} {\bibfnamefont {F.~F.}\ \bibnamefont {Balakirev}}, \bibinfo {author} {\bibfnamefont {D.}~\bibnamefont {Graf}}, \bibinfo {author} {\bibfnamefont {V.}~\bibnamefont {Prakapenka}}, \emph {et~al.},\ }\bibfield  {title} {\bibinfo {title} {Superconductivity at 250 k in lanthanum hydride under high pressures},\ }\href@noop {} {\bibfield  {journal} {\bibinfo  {journal} {Nature}\ }\textbf {\bibinfo {volume} {569}},\ \bibinfo {pages} {528} (\bibinfo {year} {2019})}\BibitemShut {NoStop}%
\bibitem [{\citenamefont {Snider}\ \emph {et~al.}(2021)\citenamefont {Snider}, \citenamefont {Dasenbrock-Gammon}, \citenamefont {McBride}, \citenamefont {Wang}, \citenamefont {Meyers}, \citenamefont {Lawler}, \citenamefont {Zurek}, \citenamefont {Salamat},\ and\ \citenamefont {Dias}}]{snider2021synthesis}%
  \BibitemOpen
  \bibfield  {author} {\bibinfo {author} {\bibfnamefont {E.}~\bibnamefont {Snider}}, \bibinfo {author} {\bibfnamefont {N.}~\bibnamefont {Dasenbrock-Gammon}}, \bibinfo {author} {\bibfnamefont {R.}~\bibnamefont {McBride}}, \bibinfo {author} {\bibfnamefont {X.}~\bibnamefont {Wang}}, \bibinfo {author} {\bibfnamefont {N.}~\bibnamefont {Meyers}}, \bibinfo {author} {\bibfnamefont {K.~V.}\ \bibnamefont {Lawler}}, \bibinfo {author} {\bibfnamefont {E.}~\bibnamefont {Zurek}}, \bibinfo {author} {\bibfnamefont {A.}~\bibnamefont {Salamat}},\ and\ \bibinfo {author} {\bibfnamefont {R.~P.}\ \bibnamefont {Dias}},\ }\bibfield  {title} {\bibinfo {title} {Synthesis of yttrium superhydride superconductor with a transition temperature up to 262 k by catalytic hydrogenation at high pressures},\ }\href@noop {} {\bibfield  {journal} {\bibinfo  {journal} {Physical review letters}\ }\textbf {\bibinfo {volume} {126}},\ \bibinfo {pages} {117003} (\bibinfo {year} {2021})}\BibitemShut {NoStop}%
\bibitem [{\citenamefont {Sukmas}\ \emph {et~al.}(2020)\citenamefont {Sukmas}, \citenamefont {Tsuppayakorn-aek}, \citenamefont {Pinsook},\ and\ \citenamefont {Bovornratanaraks}}]{sukmas2020near}%
  \BibitemOpen
  \bibfield  {author} {\bibinfo {author} {\bibfnamefont {W.}~\bibnamefont {Sukmas}}, \bibinfo {author} {\bibfnamefont {P.}~\bibnamefont {Tsuppayakorn-aek}}, \bibinfo {author} {\bibfnamefont {U.}~\bibnamefont {Pinsook}},\ and\ \bibinfo {author} {\bibfnamefont {T.}~\bibnamefont {Bovornratanaraks}},\ }\bibfield  {title} {\bibinfo {title} {Near-room-temperature superconductivity of mg/ca substituted metal hexahydride under pressure},\ }\href@noop {} {\bibfield  {journal} {\bibinfo  {journal} {Journal of Alloys and Compounds}\ }\textbf {\bibinfo {volume} {849}},\ \bibinfo {pages} {156434} (\bibinfo {year} {2020})}\BibitemShut {NoStop}%
\bibitem [{\citenamefont {Cai}\ \emph {et~al.}(2023)\citenamefont {Cai}, \citenamefont {Minkov}, \citenamefont {Sun}, \citenamefont {Kong}, \citenamefont {Sawchuk}, \citenamefont {Maiorov}, \citenamefont {Balakirev}, \citenamefont {Chariton}, \citenamefont {Prakapenka}, \citenamefont {Ma} \emph {et~al.}}]{cai2023superconductivity}%
  \BibitemOpen
  \bibfield  {author} {\bibinfo {author} {\bibfnamefont {W.}~\bibnamefont {Cai}}, \bibinfo {author} {\bibfnamefont {V.~S.}\ \bibnamefont {Minkov}}, \bibinfo {author} {\bibfnamefont {Y.}~\bibnamefont {Sun}}, \bibinfo {author} {\bibfnamefont {P.}~\bibnamefont {Kong}}, \bibinfo {author} {\bibfnamefont {K.}~\bibnamefont {Sawchuk}}, \bibinfo {author} {\bibfnamefont {B.}~\bibnamefont {Maiorov}}, \bibinfo {author} {\bibfnamefont {F.~F.}\ \bibnamefont {Balakirev}}, \bibinfo {author} {\bibfnamefont {S.}~\bibnamefont {Chariton}}, \bibinfo {author} {\bibfnamefont {V.~B.}\ \bibnamefont {Prakapenka}}, \bibinfo {author} {\bibfnamefont {Y.}~\bibnamefont {Ma}}, \emph {et~al.},\ }\bibfield  {title} {\bibinfo {title} {Superconductivity above 180 k in ca-mg ternary superhydrides at megabar pressures},\ }\href@noop {} {\bibfield  {journal} {\bibinfo  {journal} {arXiv preprint arXiv:2312.06090}\ } (\bibinfo {year} {2023})}\BibitemShut {NoStop}%
\bibitem [{\citenamefont {Baerlocher}\ and\ \citenamefont {McCusker}(2017{\natexlab{a}})}]{Baerlocher2017}%
  \BibitemOpen
  \bibfield  {author} {\bibinfo {author} {\bibfnamefont {C.}~\bibnamefont {Baerlocher}}\ and\ \bibinfo {author} {\bibfnamefont {L.~B.}\ \bibnamefont {McCusker}},\ }\href {http://www.iza-structure.org/databases/} {\bibinfo {title} {Database of zeolite structures}} (\bibinfo {year} {2017}{\natexlab{a}}),\ \bibinfo {note} {accessed: March 25, 2025}\BibitemShut {NoStop}%
\bibitem [{\citenamefont {An}\ \emph {et~al.}(2024)\citenamefont {An}, \citenamefont {Duan}, \citenamefont {Zhang}, \citenamefont {Jiang}, \citenamefont {Ma}, \citenamefont {Huo}, \citenamefont {Song},\ and\ \citenamefont {Cui}}]{an2024type}%
  \BibitemOpen
  \bibfield  {author} {\bibinfo {author} {\bibfnamefont {D.}~\bibnamefont {An}}, \bibinfo {author} {\bibfnamefont {D.}~\bibnamefont {Duan}}, \bibinfo {author} {\bibfnamefont {Z.}~\bibnamefont {Zhang}}, \bibinfo {author} {\bibfnamefont {Q.}~\bibnamefont {Jiang}}, \bibinfo {author} {\bibfnamefont {T.}~\bibnamefont {Ma}}, \bibinfo {author} {\bibfnamefont {Z.}~\bibnamefont {Huo}}, \bibinfo {author} {\bibfnamefont {H.}~\bibnamefont {Song}},\ and\ \bibinfo {author} {\bibfnamefont {T.}~\bibnamefont {Cui}},\ }\bibfield  {title} {\bibinfo {title} {Type-i clathrate calcium hydride and its hydrogen-vacancy structures at high pressure},\ }\href@noop {} {\bibfield  {journal} {\bibinfo  {journal} {Physical Review B}\ }\textbf {\bibinfo {volume} {110}},\ \bibinfo {pages} {054505} (\bibinfo {year} {2024})}\BibitemShut {NoStop}%
\bibitem [{\citenamefont {Jiang}\ \emph {et~al.}(2025)\citenamefont {Jiang}, \citenamefont {Luo}, \citenamefont {Sun}, \citenamefont {Zhong}, \citenamefont {Lv}, \citenamefont {Xie}, \citenamefont {Ma},\ and\ \citenamefont {Liu}}]{jiang2025data}%
  \BibitemOpen
  \bibfield  {author} {\bibinfo {author} {\bibfnamefont {B.}~\bibnamefont {Jiang}}, \bibinfo {author} {\bibfnamefont {X.}~\bibnamefont {Luo}}, \bibinfo {author} {\bibfnamefont {Y.}~\bibnamefont {Sun}}, \bibinfo {author} {\bibfnamefont {X.}~\bibnamefont {Zhong}}, \bibinfo {author} {\bibfnamefont {J.}~\bibnamefont {Lv}}, \bibinfo {author} {\bibfnamefont {Y.}~\bibnamefont {Xie}}, \bibinfo {author} {\bibfnamefont {Y.}~\bibnamefont {Ma}},\ and\ \bibinfo {author} {\bibfnamefont {H.}~\bibnamefont {Liu}},\ }\bibfield  {title} {\bibinfo {title} {Data-driven search for high-temperature superconductors in ternary hydrides under pressure},\ }\href@noop {} {\bibfield  {journal} {\bibinfo  {journal} {Physical Review B}\ }\textbf {\bibinfo {volume} {111}},\ \bibinfo {pages} {054505} (\bibinfo {year} {2025})}\BibitemShut {NoStop}%
\bibitem [{\citenamefont {Baerlocher}\ and\ \citenamefont {McCusker}(2017{\natexlab{b}})}]{baerlocher2017database}%
  \BibitemOpen
  \bibfield  {author} {\bibinfo {author} {\bibfnamefont {C.}~\bibnamefont {Baerlocher}}\ and\ \bibinfo {author} {\bibfnamefont {L.~B.}\ \bibnamefont {McCusker}},\ }\href@noop {} {\bibinfo {title} {Database of zeolite structures}},\ \bibinfo {howpublished} {\url{http://www.iza-structure.org/databases/}} (\bibinfo {year} {2017}{\natexlab{b}})\BibitemShut {NoStop}%
\bibitem [{\citenamefont {{Reticular Chemistry Structure Resource}}()}]{rcsr_database}%
  \BibitemOpen
  \bibfield  {author} {\bibinfo {author} {\bibnamefont {{Reticular Chemistry Structure Resource}}},\ }\href@noop {} {\bibinfo {title} {Reticular chemistry structure resource}},\ \bibinfo {howpublished} {\url{http://rcsr.anu.edu.au/}}\BibitemShut {NoStop}%
\bibitem [{\citenamefont {He}\ \emph {et~al.}(2024)\citenamefont {He}, \citenamefont {Zhao}, \citenamefont {Xie}, \citenamefont {Hermann}, \citenamefont {Hemley}, \citenamefont {Liu},\ and\ \citenamefont {Ma}}]{he2024predicted}%
  \BibitemOpen
  \bibfield  {author} {\bibinfo {author} {\bibfnamefont {X.-L.}\ \bibnamefont {He}}, \bibinfo {author} {\bibfnamefont {W.}~\bibnamefont {Zhao}}, \bibinfo {author} {\bibfnamefont {Y.}~\bibnamefont {Xie}}, \bibinfo {author} {\bibfnamefont {A.}~\bibnamefont {Hermann}}, \bibinfo {author} {\bibfnamefont {R.~J.}\ \bibnamefont {Hemley}}, \bibinfo {author} {\bibfnamefont {H.}~\bibnamefont {Liu}},\ and\ \bibinfo {author} {\bibfnamefont {Y.}~\bibnamefont {Ma}},\ }\bibfield  {title} {\bibinfo {title} {Predicted hot superconductivity in lasc2h24 under pressure},\ }\href@noop {} {\bibfield  {journal} {\bibinfo  {journal} {Proceedings of the National Academy of Sciences}\ }\textbf {\bibinfo {volume} {121}},\ \bibinfo {pages} {e2401840121} (\bibinfo {year} {2024})}\BibitemShut {NoStop}%
\bibitem [{\citenamefont {Yu}\ \emph {et~al.}(2017)\citenamefont {Yu}, \citenamefont {Young}, \citenamefont {Wu}, \citenamefont {Zhang}, \citenamefont {Rondinelli},\ and\ \citenamefont {Halasyamani}}]{yu2017m4mg4}%
  \BibitemOpen
  \bibfield  {author} {\bibinfo {author} {\bibfnamefont {H.}~\bibnamefont {Yu}}, \bibinfo {author} {\bibfnamefont {J.}~\bibnamefont {Young}}, \bibinfo {author} {\bibfnamefont {H.}~\bibnamefont {Wu}}, \bibinfo {author} {\bibfnamefont {W.}~\bibnamefont {Zhang}}, \bibinfo {author} {\bibfnamefont {J.~M.}\ \bibnamefont {Rondinelli}},\ and\ \bibinfo {author} {\bibfnamefont {P.~S.}\ \bibnamefont {Halasyamani}},\ }\bibfield  {title} {\bibinfo {title} {M4mg4 (p2o7) 3 (m= k, rb): structural engineering of pyrophosphates for nonlinear optical applications},\ }\href@noop {} {\bibfield  {journal} {\bibinfo  {journal} {Chemistry of Materials}\ }\textbf {\bibinfo {volume} {29}},\ \bibinfo {pages} {1845} (\bibinfo {year} {2017})}\BibitemShut {NoStop}%
\bibitem [{\citenamefont {Li}\ \emph {et~al.}(2016)\citenamefont {Li}, \citenamefont {Wang}, \citenamefont {Lei}, \citenamefont {Han}, \citenamefont {Yang}, \citenamefont {Poeppelmeier},\ and\ \citenamefont {Pan}}]{li2016new}%
  \BibitemOpen
  \bibfield  {author} {\bibinfo {author} {\bibfnamefont {L.}~\bibnamefont {Li}}, \bibinfo {author} {\bibfnamefont {Y.}~\bibnamefont {Wang}}, \bibinfo {author} {\bibfnamefont {B.-H.}\ \bibnamefont {Lei}}, \bibinfo {author} {\bibfnamefont {S.}~\bibnamefont {Han}}, \bibinfo {author} {\bibfnamefont {Z.}~\bibnamefont {Yang}}, \bibinfo {author} {\bibfnamefont {K.~R.}\ \bibnamefont {Poeppelmeier}},\ and\ \bibinfo {author} {\bibfnamefont {S.}~\bibnamefont {Pan}},\ }\bibfield  {title} {\bibinfo {title} {A new deep-ultraviolet transparent orthophosphate lics2po4 with large second harmonic generation response},\ }\href@noop {} {\bibfield  {journal} {\bibinfo  {journal} {Journal of the American Chemical Society}\ }\textbf {\bibinfo {volume} {138}},\ \bibinfo {pages} {9101} (\bibinfo {year} {2016})}\BibitemShut {NoStop}%
\bibitem [{\citenamefont {Xie}\ \emph {et~al.}(2022)\citenamefont {Xie}, \citenamefont {Tudi},\ and\ \citenamefont {Oganov}}]{xie2022pno}%
  \BibitemOpen
  \bibfield  {author} {\bibinfo {author} {\bibfnamefont {C.}~\bibnamefont {Xie}}, \bibinfo {author} {\bibfnamefont {A.}~\bibnamefont {Tudi}},\ and\ \bibinfo {author} {\bibfnamefont {A.~R.}\ \bibnamefont {Oganov}},\ }\bibfield  {title} {\bibinfo {title} {Pno: a promising deep-uv nonlinear optical material with the largest second harmonic generation effect},\ }\href@noop {} {\bibfield  {journal} {\bibinfo  {journal} {Chemical Communications}\ }\textbf {\bibinfo {volume} {58}},\ \bibinfo {pages} {12491} (\bibinfo {year} {2022})}\BibitemShut {NoStop}%
\bibitem [{\citenamefont {Lyakhov}\ \emph {et~al.}(2010)\citenamefont {Lyakhov}, \citenamefont {Oganov},\ and\ \citenamefont {Valle}}]{lyakhov2010predict}%
  \BibitemOpen
  \bibfield  {author} {\bibinfo {author} {\bibfnamefont {A.~O.}\ \bibnamefont {Lyakhov}}, \bibinfo {author} {\bibfnamefont {A.~R.}\ \bibnamefont {Oganov}},\ and\ \bibinfo {author} {\bibfnamefont {M.}~\bibnamefont {Valle}},\ }\bibfield  {title} {\bibinfo {title} {How to predict very large and complex crystal structures},\ }\href@noop {} {\bibfield  {journal} {\bibinfo  {journal} {Computer Physics Communications}\ }\textbf {\bibinfo {volume} {181}},\ \bibinfo {pages} {1623} (\bibinfo {year} {2010})}\BibitemShut {NoStop}%
\bibitem [{\citenamefont {Antunes}\ \emph {et~al.}(2024)\citenamefont {Antunes}, \citenamefont {Butler},\ and\ \citenamefont {Grau-Crespo}}]{antunes2024crystal}%
  \BibitemOpen
  \bibfield  {author} {\bibinfo {author} {\bibfnamefont {L.~M.}\ \bibnamefont {Antunes}}, \bibinfo {author} {\bibfnamefont {K.~T.}\ \bibnamefont {Butler}},\ and\ \bibinfo {author} {\bibfnamefont {R.}~\bibnamefont {Grau-Crespo}},\ }\bibfield  {title} {\bibinfo {title} {Crystal structure generation with autoregressive large language modeling},\ }\href@noop {} {\bibfield  {journal} {\bibinfo  {journal} {Nature Communications}\ }\textbf {\bibinfo {volume} {15}},\ \bibinfo {pages} {1} (\bibinfo {year} {2024})}\BibitemShut {NoStop}%
\bibitem [{\citenamefont {Luo}\ \emph {et~al.}(2024)\citenamefont {Luo}, \citenamefont {Wang}, \citenamefont {Gao}, \citenamefont {Lv}, \citenamefont {Wang}, \citenamefont {Chen},\ and\ \citenamefont {Ma}}]{luo2024deep}%
  \BibitemOpen
  \bibfield  {author} {\bibinfo {author} {\bibfnamefont {X.}~\bibnamefont {Luo}}, \bibinfo {author} {\bibfnamefont {Z.}~\bibnamefont {Wang}}, \bibinfo {author} {\bibfnamefont {P.}~\bibnamefont {Gao}}, \bibinfo {author} {\bibfnamefont {J.}~\bibnamefont {Lv}}, \bibinfo {author} {\bibfnamefont {Y.}~\bibnamefont {Wang}}, \bibinfo {author} {\bibfnamefont {C.}~\bibnamefont {Chen}},\ and\ \bibinfo {author} {\bibfnamefont {Y.}~\bibnamefont {Ma}},\ }\bibfield  {title} {\bibinfo {title} {Deep learning generative model for crystal structure prediction},\ }\href@noop {} {\bibfield  {journal} {\bibinfo  {journal} {npj Computational Materials}\ }\textbf {\bibinfo {volume} {10}},\ \bibinfo {pages} {254} (\bibinfo {year} {2024})}\BibitemShut {NoStop}%
\bibitem [{\citenamefont {Xie}\ \emph {et~al.}(2021)\citenamefont {Xie}, \citenamefont {Fu}, \citenamefont {Ganea}, \citenamefont {Barzilay},\ and\ \citenamefont {Jaakkola}}]{xie2021crystal}%
  \BibitemOpen
  \bibfield  {author} {\bibinfo {author} {\bibfnamefont {T.}~\bibnamefont {Xie}}, \bibinfo {author} {\bibfnamefont {X.}~\bibnamefont {Fu}}, \bibinfo {author} {\bibfnamefont {O.-E.}\ \bibnamefont {Ganea}}, \bibinfo {author} {\bibfnamefont {R.}~\bibnamefont {Barzilay}},\ and\ \bibinfo {author} {\bibfnamefont {T.}~\bibnamefont {Jaakkola}},\ }\bibfield  {title} {\bibinfo {title} {Crystal diffusion variational autoencoder for periodic material generation},\ }\href@noop {} {\bibfield  {journal} {\bibinfo  {journal} {arXiv preprint arXiv:2110.06197}\ } (\bibinfo {year} {2021})}\BibitemShut {NoStop}%
\bibitem [{\citenamefont {Vasylenko}\ \emph {et~al.}(2024)\citenamefont {Vasylenko}, \citenamefont {Asher}, \citenamefont {Collins}, \citenamefont {Gaultois}, \citenamefont {Darling}, \citenamefont {Dyer},\ and\ \citenamefont {Rosseinsky}}]{vasylenko2024inferring}%
  \BibitemOpen
  \bibfield  {author} {\bibinfo {author} {\bibfnamefont {A.}~\bibnamefont {Vasylenko}}, \bibinfo {author} {\bibfnamefont {B.~M.}\ \bibnamefont {Asher}}, \bibinfo {author} {\bibfnamefont {C.~M.}\ \bibnamefont {Collins}}, \bibinfo {author} {\bibfnamefont {M.~W.}\ \bibnamefont {Gaultois}}, \bibinfo {author} {\bibfnamefont {G.~R.}\ \bibnamefont {Darling}}, \bibinfo {author} {\bibfnamefont {M.~S.}\ \bibnamefont {Dyer}},\ and\ \bibinfo {author} {\bibfnamefont {M.~J.}\ \bibnamefont {Rosseinsky}},\ }\bibfield  {title} {\bibinfo {title} {Inferring energy--composition relationships with bayesian optimization enhances exploration of inorganic materials},\ }\href@noop {} {\bibfield  {journal} {\bibinfo  {journal} {The Journal of chemical physics}\ }\textbf {\bibinfo {volume} {160}} (\bibinfo {year} {2024})}\BibitemShut {NoStop}%
\bibitem [{\citenamefont {Xu}\ \emph {et~al.}(2019)\citenamefont {Xu}, \citenamefont {Wang}, \citenamefont {Xue}, \citenamefont {Shao}, \citenamefont {Gao}, \citenamefont {Lv}, \citenamefont {Wang},\ and\ \citenamefont {Ma}}]{xu2019ab}%
  \BibitemOpen
  \bibfield  {author} {\bibinfo {author} {\bibfnamefont {Q.}~\bibnamefont {Xu}}, \bibinfo {author} {\bibfnamefont {S.}~\bibnamefont {Wang}}, \bibinfo {author} {\bibfnamefont {L.}~\bibnamefont {Xue}}, \bibinfo {author} {\bibfnamefont {X.}~\bibnamefont {Shao}}, \bibinfo {author} {\bibfnamefont {P.}~\bibnamefont {Gao}}, \bibinfo {author} {\bibfnamefont {J.}~\bibnamefont {Lv}}, \bibinfo {author} {\bibfnamefont {Y.}~\bibnamefont {Wang}},\ and\ \bibinfo {author} {\bibfnamefont {Y.}~\bibnamefont {Ma}},\ }\bibfield  {title} {\bibinfo {title} {Ab initio electronic structure calculations using a real-space chebyshev-filtered subspace iteration method},\ }\href@noop {} {\bibfield  {journal} {\bibinfo  {journal} {Journal of Physics: Condensed Matter}\ }\textbf {\bibinfo {volume} {31}},\ \bibinfo {pages} {455901} (\bibinfo {year} {2019})}\BibitemShut {NoStop}%
\bibitem [{\citenamefont {Loshchilov}\ and\ \citenamefont {Hutter}(2017)}]{loshchilov2017decoupled}%
  \BibitemOpen
  \bibfield  {author} {\bibinfo {author} {\bibfnamefont {I.}~\bibnamefont {Loshchilov}}\ and\ \bibinfo {author} {\bibfnamefont {F.}~\bibnamefont {Hutter}},\ }\bibfield  {title} {\bibinfo {title} {Decoupled weight decay regularization},\ }\href@noop {} {\bibfield  {journal} {\bibinfo  {journal} {arXiv preprint arXiv:1711.05101}\ } (\bibinfo {year} {2017})}\BibitemShut {NoStop}%
\bibitem [{\citenamefont {Perdew}\ \emph {et~al.}(1996)\citenamefont {Perdew}, \citenamefont {Burke},\ and\ \citenamefont {Ernzerhof}}]{perdew1996generalized}%
  \BibitemOpen
  \bibfield  {author} {\bibinfo {author} {\bibfnamefont {J.~P.}\ \bibnamefont {Perdew}}, \bibinfo {author} {\bibfnamefont {K.}~\bibnamefont {Burke}},\ and\ \bibinfo {author} {\bibfnamefont {M.}~\bibnamefont {Ernzerhof}},\ }\bibfield  {title} {\bibinfo {title} {Generalized gradient approximation made simple},\ }\href@noop {} {\bibfield  {journal} {\bibinfo  {journal} {Physical review letters}\ }\textbf {\bibinfo {volume} {77}},\ \bibinfo {pages} {3865} (\bibinfo {year} {1996})}\BibitemShut {NoStop}%
\bibitem [{\citenamefont {Hamann}(2013)}]{hamann2013optimized}%
  \BibitemOpen
  \bibfield  {author} {\bibinfo {author} {\bibfnamefont {D.}~\bibnamefont {Hamann}},\ }\bibfield  {title} {\bibinfo {title} {Optimized norm-conserving vanderbilt pseudopotentials},\ }\href@noop {} {\bibfield  {journal} {\bibinfo  {journal} {Physical Review B—Condensed Matter and Materials Physics}\ }\textbf {\bibinfo {volume} {88}},\ \bibinfo {pages} {085117} (\bibinfo {year} {2013})}\BibitemShut {NoStop}%
\bibitem [{\citenamefont {Van~Setten}\ \emph {et~al.}(2018)\citenamefont {Van~Setten}, \citenamefont {Giantomassi}, \citenamefont {Bousquet}, \citenamefont {Verstraete}, \citenamefont {Hamann}, \citenamefont {Gonze},\ and\ \citenamefont {Rignanese}}]{van2018pseudodojo}%
  \BibitemOpen
  \bibfield  {author} {\bibinfo {author} {\bibfnamefont {M.~J.}\ \bibnamefont {Van~Setten}}, \bibinfo {author} {\bibfnamefont {M.}~\bibnamefont {Giantomassi}}, \bibinfo {author} {\bibfnamefont {E.}~\bibnamefont {Bousquet}}, \bibinfo {author} {\bibfnamefont {M.~J.}\ \bibnamefont {Verstraete}}, \bibinfo {author} {\bibfnamefont {D.~R.}\ \bibnamefont {Hamann}}, \bibinfo {author} {\bibfnamefont {X.}~\bibnamefont {Gonze}},\ and\ \bibinfo {author} {\bibfnamefont {G.-M.}\ \bibnamefont {Rignanese}},\ }\bibfield  {title} {\bibinfo {title} {The pseudodojo: Training and grading a 85 element optimized norm-conserving pseudopotential table},\ }\href@noop {} {\bibfield  {journal} {\bibinfo  {journal} {Computer Physics Communications}\ }\textbf {\bibinfo {volume} {226}},\ \bibinfo {pages} {39} (\bibinfo {year} {2018})}\BibitemShut {NoStop}%
\bibitem [{\citenamefont {Nocedal}\ and\ \citenamefont {Wright}(2006)}]{nocedal2006numerical}%
  \BibitemOpen
  \bibfield  {author} {\bibinfo {author} {\bibfnamefont {J.}~\bibnamefont {Nocedal}}\ and\ \bibinfo {author} {\bibfnamefont {S.~J.}\ \bibnamefont {Wright}},\ }\href@noop {} {\emph {\bibinfo {title} {Numerical Optimization}}},\ \bibinfo {edition} {2nd}\ ed.\ (\bibinfo  {publisher} {Springer},\ \bibinfo {address} {New York, NY, USA},\ \bibinfo {year} {2006})\BibitemShut {NoStop}%
\end{thebibliography}%

\end{document}